\documentclass[10]{article}

\usepackage{amsmath,graphicx,bbm,amssymb}
\usepackage{caption}
\usepackage{color}
\usepackage{tikz}
\usetikzlibrary{shapes.multipart}
\usetikzlibrary{shapes.misc, positioning}
\usepackage{fancyhdr}
\usepackage{colortbl}
\usepackage{marvosym}
\usepackage[hmargin=1.5cm, top=2cm, bottom=2cm]{geometry}
\usepackage[colorlinks=true,linkcolor=blue,citecolor=blue,urlcolor=blue]{hyperref}
\usepackage[export]{adjustbox}
\usetikzlibrary{arrows} 
\newcommand{\norme}[1]{\left\vert\left\vert #1 \right\vert\right\vert}
\newcommand{\para}[1]{\left(#1\right)}

\newcommand{\rz}{r_0}

\newcommand{\cnh}{C_n^2(h)}

\newcommand{\rbb}{\boldsymbol{r}}

\newcommand{\rhob}{\boldsymbol{\rho}}

\newcommand{\otf}[1]{\text{OTF}_{#1}}
\newcommand{\atf}[1]{\text{ATF}_{#1}}

\newcommand{\wPSF}{\widehat{\text{PSF}}}

\newcommand{\pup}{\mathcal{P}}

\newcolumntype{P}[1]{>{\centering\arraybackslash}p{#1}}
\usepackage{titling}
\usepackage[affil-it]{authblk}

\title{\LARGE Focal plane ${\cnh}$ profiling based on single conjugate adaptive optics compensated images.}

\author{
	O. Beltramo-Martin,$^{1}$\thanks{E-mail: olivier.beltramo-martin@lam.fr}
	C.M. Correia,$^{1}$	
	B. Neichel,$^{1}$        
	T. Fusco,$^{1,2}$    
	}
	
	\affil{
	\small $^{1}$Aix Marseille Univ., CNRS, LAM, Laboratoire d'Astrophysique de Marseille, 38 rue F. Joliot-Curie, 13388 Marseille, France\\
	$^{2}$ONERA, The French Aerospace Laboratory  BP. 72, F-92322 Chatillon Cedex, France\\	
}

\date{}

\begin{document}
	\label{firstpage}
	\maketitle
	
	\begin{abstract}	
    \textcolor{black}{Knowledge of the atmospheric turbulence in the telescope line-of-sight is crucial for wide-field observations assisted by adaptive optics (AO)}, for which the \textcolor{black}{Point Spread Function} (PSF) becomes strongly elongated due to the anisoplanatism effect. This one must be modelled accurately to extrapolate the PSF anywhere across the Field of view (FOV) and improve the science exploitation. However, anisoplanatism is a function of the $\cnh$ profile, that is not directly accessible from single conjugate AO telemetry. \textcolor{black}{One may rely on external profilers, but recent studies have highlighted more than 10\% of discrepancies with AO internal measurements, while we aim better than 1\% of accuracy for PSF modelling.}
To tackle this existing limitation, we present the Focal plane profiling (FPP), as a
$\cnh$ profiling method that relies on post-AO focal plane images. We demonstrate such an approach complies with a 1\%-level of accuracy on the $\cnh$ estimation and establish how this accuracy varies regarding the calibration stars magnitudes and positions in the field. We highlight that  photometry and astrometry errors due to PSF mis-modelling reaches respectively 1\% and 50$\mu$as using FPP on a Keck baseline, with a preliminary calibration using a star of magnitude H=14 at 20". We also validate this concept using Canada’s NRC-Herzberg HENOS testbed images in comparing FPP retrieval with alternative $\cnh$ measurements on HeNOS. The FPP approach allows to profile the $\cnh$ using the SCAO systems and improve significantly the PSF characterisation. Such a methodology is also ELT-size compliant and will be extrapolated to tomographic systems in a near future.
	\end{abstract}


	\section{Introduction}
	
	This paper focuses on improving the AO PSF characterisation in a wide-field by retrieving the distribution of atmospheric turbulence along altitude it depends on, referred as the $\cnh$ profile. We consider Single Conjugate Adaptive Optics (SCAO) system-assisted observations; \textcolor{black}{the correction provided by AO is optimal in the Guide star~(GS) direction, that can be either natural~(NGS) or artificial using laser~(LGS)}, but degrades across the field on account of the anisoplanatism effect~\cite{Fried1982}. This \textcolor{black}{latter} results from the spatial decorrelation of the incoming electric field phase that propagates through the atmosphere. The way this decorrelation occurs is a direct function of the $\cnh$ and outer scale profiles. As a consequence, anisoplanatism broaden the PSF and induces a spatial variation of the PSF morphology across the field.\\
	
	However, the PSF model is one of the key limitations in the current exploitation of images of crowded-field stellar populations~\cite{Fritz2010,Yelda2010,Shodel2010} that are affected by anisoplanatism. To strengthen the data-processing outcome, we propose to improve the anisoplanatism characterisation. We have established a complete and general anisoplanatism model in~\cite{BeltramoMartin2018} as a function of the input $\cnh$ profile. 
	This information is not accessible from the AO telemetry for SCAO systems, though. Dedicated instruments exist to monitor this profile~\cite{Osborn2015,Butterley2006,Wilson2002}, but they aim at characterising the observation site in terms of atmosphere quality and do not observe in the telescope line of sight. Consequently, their estimated profile deviate up to at least 10~\%~\cite{Ono2017}, from AO telemetry-based approaches available for multi-GS AO systems~\cite{Helin2018,Guesalaga2017,Laidlaw2016,Martin2016L3S,Neichel2014}. \textcolor{black}{However, according to \cite{BeltramoMartin2018}, 10\% of error on the $\cnh$ estimation may degrade the photometry and astrometry determination up to respectively 3\% and 300 $\mu$as in a 20"-FOV, while we are seeking to reach better than 1\% and 150 $\mu$as.}  \\
	
	We propose in this paper to rely on the SCAO-compensated PSFs available across the field. \textcolor{black}{We focus on SCAO systems that do not permit to identify the $\cnh$ from the telemetry and would benefit to have an internal images-based facility to retrieve the profile, either for post-processing or real-time application. The methodology we present can be extended to multi-GS systems, but at the cost of a larger numerical complexity to include the AO system control specificity, such as the tomographic reconstruction step or the optimal fitting in multi-conjugated AO. Before considering such systems, we will address the ground-layer AO case and compare our present approach to telemetry-based $\cnh$ estimation.}\\

	Spatial variations of the AO PSF encodes the real $\cnh$ that affects images and the one that we want to determine. To extract the profile from images, we have developed the Focal plane profiling~(FPP) method as a non-linear least-square minimisation procedure that adjust a $\cnh$-dependent PSF model to match a collection of observations and deliver a joint estimation of the PSF model and $\cnh$ profile. If the model is not consistent with real atmosphere statistics, because we consider a profile over too few bins for instance, the minimisation process allows to mitigate these errors when extrapolating the PSF in the field. At the contrary, if we feed this inaccurate model with a wrong $\cnh$ without any feedback from a real PSF, we risk to amplify the error propagation and \textcolor{black}{degrade} the PSF extrapolation.\\
	
	We describe the FPP algorithm in Sect.~\ref{S:FPP}. Sect.~\ref{S:simu} is dedicated to FPP performance assessment; we illustrate that the FPP allows to retrieve the $\cnh$ at a 1\%-level accuracy when bright stars are available. We present an application to PSF extrapolation on simulated images of NIRC2 \cite{McLean2000} at Keck II and evaluate what are the conditions in terms of calibration stars magnitudes and field location to decrease errors provoked by PSF-model indetermination on photometry and astrometry, down to respectively 1\% and 50$\mu$as. We make a step further in applying the FPP to Canada's NRC-Herzberg HENOS testbed images and compare successfully the FPP-retrieved $\cnh$ to existing measurements.

	\section{Focal plane profiling}
	\label{S:FPP}
	
	The FWHM of a seeing-limited PSF is known to be fully related to atmospheric turbulence properties, such as the seeing and the outer scale. Thus, these parameters can be extracted directly from observed PSFs as it has already been investigated and validated~\cite{Martinez2010}. 
	
	In this paper, we extend this methodology to characterise the entire $\cnh$ profile based on AO-compensated PSF. FPP is designed to exploit anisoplanatism patterns observed on off-axis PSFs, i.e. not in the guide star location, to retrieve the $\cnh$ profile as really seen by the AO system. It relies on a PSF model~\cite{BeltramoMartin2018} regarding angular variations across the field due to anisoplanatism. Furthermore, to inverse the problem, it performs iteratively a joint estimation of the PSF and the input parameters, based on a non-linear least-square minimisation yielded by a Levenberg-Marquardt algorithm.

	\subsection{Direct problem}
	
	Let $\rhob$, $\lambda$ and $\theta$ be respectively the coordinates vector in the pupil, the imaging wavelength and the angular position in the sky. We define $\otf{}(\rhob/\lambda,\cnh,\theta)$ as an AO-compensated estimated Optical transfer function~(OTF) in the direction $\theta$ in the field and is derived as follows
	\begin{equation}\label{E:model}
		\otf{}(\rhob/\lambda,\cnh,\theta) =  \otf{0}(\rhob/\lambda) \cdot \atf{}(\rhob/\lambda,\cnh,\theta) 
	\end{equation}
	where $\otf{0}$ is the OTF in the reference direction, that can be either a real observation or a model delivered by PSF reconstruction~\cite{Veran1997} for instance. In the rest of this paper, we will assume to perfectly know $ \otf{0}$. Then, \textcolor{black}{the PSF in the field direction $\theta$} is given by the Fourier transform of $\otf{}(\rhob/\lambda,\cnh,\theta)$.\\
	
	ATF defined in Eq.~\ref{E:model} is the Anisoplanatism transfer function as introduced by~\cite{Fusco2000} and derives as
	\begin{equation} \label{E:atf}
		\begin{aligned}
			&\text{ATF}(\rhob/\lambda,\cnh,\theta)  =\\
			& \dfrac{\iint_\pup \pup(\rbb)\pup(\rbb+\rhob)\exp\para{-0.5\times D_\Delta(\rbb,\rhob,\cnh,\theta)}\boldsymbol{dr}}{\iint_\pup \pup(\rbb)\pup(\rbb+\rhob)\boldsymbol{dr}}
		\end{aligned}
	\end{equation}
	where $\mathcal{P}$ is the pupil function, $\rbb/\rhob$ are respectively the location/separation vector in the pupil and $D_\Delta$ the anisoplanatic phase structure function. \textcolor{black}{$D_\Delta(\rbb, \rhob, \cnh, \theta)$ characterises the spatial decorrelation of two wavefront coming from two stars angularly separated by $\theta$ in the field. Through the ATF calculation in Eq.~\ref{E:atf}, it sharpens the angular frequencies support, i.e. $\otf{0}$, that elongates the PSF towards the GS direction. At a field position $\theta$, the PSF broadening is fully determined by $D_\Delta(\rbb, \rhob, \cnh, \theta)$, which is a function of the $\cnh$ as detailed in the literature \cite{BeltramoMartin2018,Flicker2008,Fusco2000,Britton2006,Whiteley1998,Tyler1994}. Note that the ATF should not be pupil model-dependent regarding Eq.~\ref{E:atf} that introduces the normalisation by the diffraction OTF.  The real telescope pupil filtering is carried by $\otf{0}$.}

	\subsection{Problem inversion}
	
\begin{figure*} 
		\centering
		\includegraphics[width = 18cm]{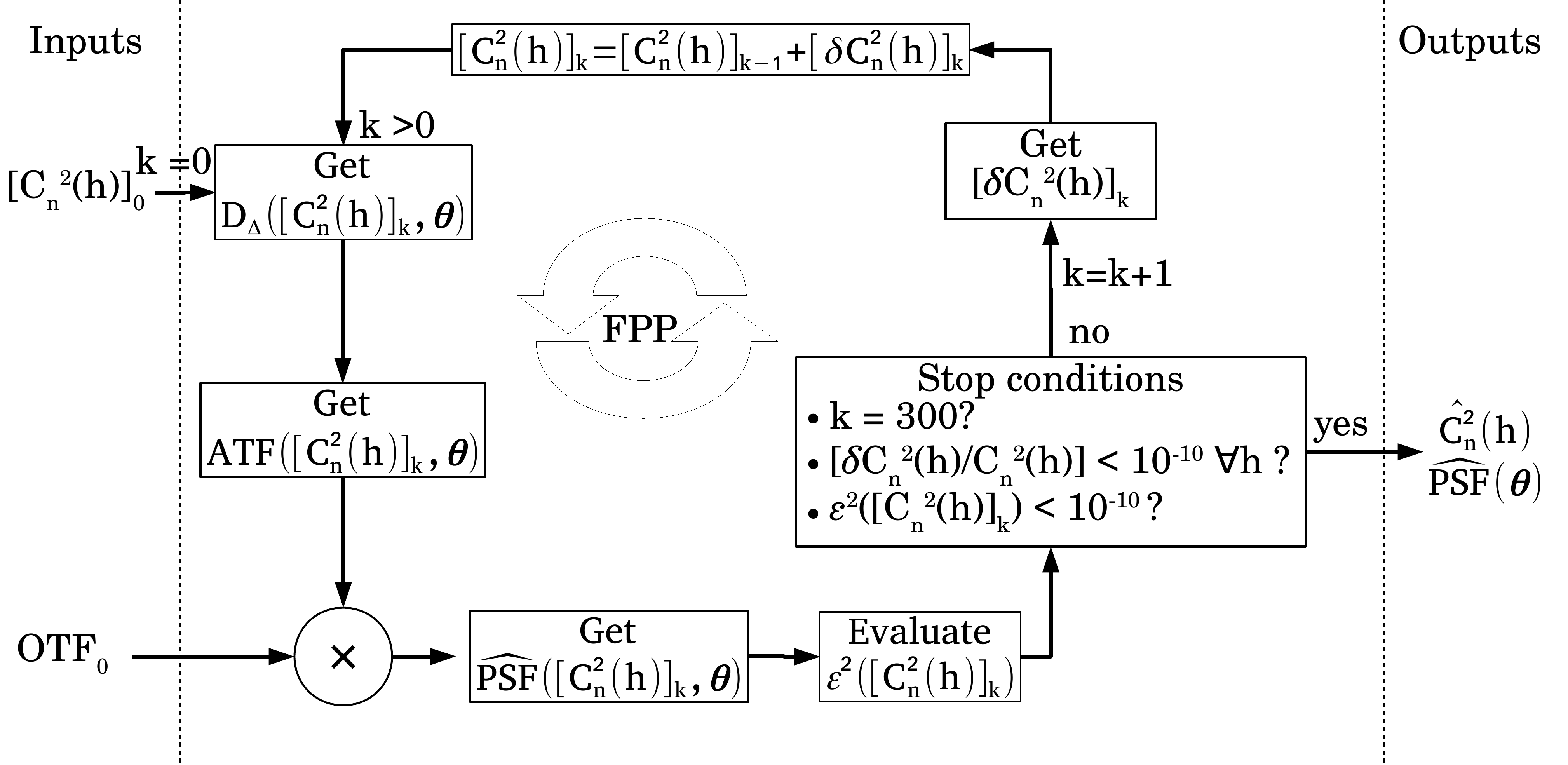}
		\caption{Block diagram representing the FPP architecture. I must be provided by a reference OTF ($\otf{0}$) and an initial guess on the $\cnh$. We have chosen a flat initial profile whose the integral corresponds to $\rz^{-5/3}$.}
		\label{F:FPP}
	\end{figure*}

	Eq.~\ref{E:atf}, highlights that the PSF model is highly non-linear regarding the inputs. The approach to inverse the problem we have chosen consists in least-squares minimising iteratively a criterion using a Levenberg-Marquardt algorithm. For $n_\text{psf}$ observations in the field, FPP minimises the following cost function
	\begin{equation}
		\varepsilon^2(\cnh) = \sum_{i=1}^{n_\text{psf}} \norme{\para{\text{PSF}(\boldsymbol{\alpha},\theta_i) - \wPSF(\boldsymbol{\alpha},\theta_i,\cnh)}}^2_{\mathcal{L}_2},
		\label{E:criteria}
	\end{equation}
	where $\text{PSF}(\boldsymbol{\alpha},\theta_i)$ and $\wPSF(\boldsymbol{\alpha},\theta_i,\cnh)$ are respectively the observed and modelled PSF in the field direction $\theta_i$ as function of the angular separation vector $\boldsymbol{\alpha}$ in the focal plane. We report in Fig.\ref{F:FPP} the FPP architecture as a block diagram.\\
	
	\textcolor{black}{It starts from an initial guess on the profile, that is chosen flat but with an integral that corresponds to $\rz^{-5/3}$. This constrain is quite easily reached thanks to the AO telemetry that permits a $\rz$ estimation within 10\% of accuracy. FPP derives the OTF in the field directions given by the vector $\boldsymbol{\theta}$ using Eq.~\ref{E:model}, where $\otf{0}$ is an input of the problem. Then, we do zero-pad and interpolate the OTF and compute its Fourier transform to get the PSF with the desired pixel-scale and FOV, before scaling it to set its flux to the observation's. From the concatenation of the modelled PSFs given at any element of $\boldsymbol{\theta}$, FPP calculates the criterion in Eq.~\ref{E:criteria} and evaluates the trueness of the stop conditions; either the criterion or the relative increment on the $\cnh$ reaches $10^{-10}$ for any bin, or when the iteration number meets 300, that is a reasonable empirical evaluation. In the situation the algorithm ought to continue getting through the iterative loop, it updates the $\cnh$ value from the empirical model gradient and update the PSF model until reaching the stop conditions. When these ones met, the FPP delivers the PSF model at any field position given by $\boldsymbol{\theta}$ and the corresponding $\cnh$ estimation. }\\

    \textcolor{black}{We could define an OTF-based criterion; we can select a Cartesian area in the PSF by applying a sinc filter to the OTF, to remove the PSF wings contribution for instance, that are not sensitive to the $\cnh$ distribution but only the integrated value. 
    Nevertheless, the anisoplanatism enlarges the PSF and does sharpen the OTF, though. Consequently, a strong anisoplanatism effect narrows the OTF and decrease the number of useful pixels to be model-fitted, which yields a sensitivity loss to $\cnh$.
    On top of that, an OTF-based fitting would potentially permit to deal more efficiently with white noise, but the processing of NIRC2 images, we will test the method on, has shown that the noise can be quite spatially correlated and contaminates more than the central OTF pixel. It is not consequently necessarily straightforward to mitigate the noise contribution in an OTF-based criterion with spatially correlated noise.
    So far the PSF-based criterion appeared to be more convenient for us in a first implementation of the method and also more intelligible for the community of potential users, but we plan to improve the FPP robustness and efficiency in the future. We will particularly figure out whether an OTF-based criterion may help to go to this direction.}\\

\begin{figure}
		\centering
		\includegraphics[width=8.8cm]{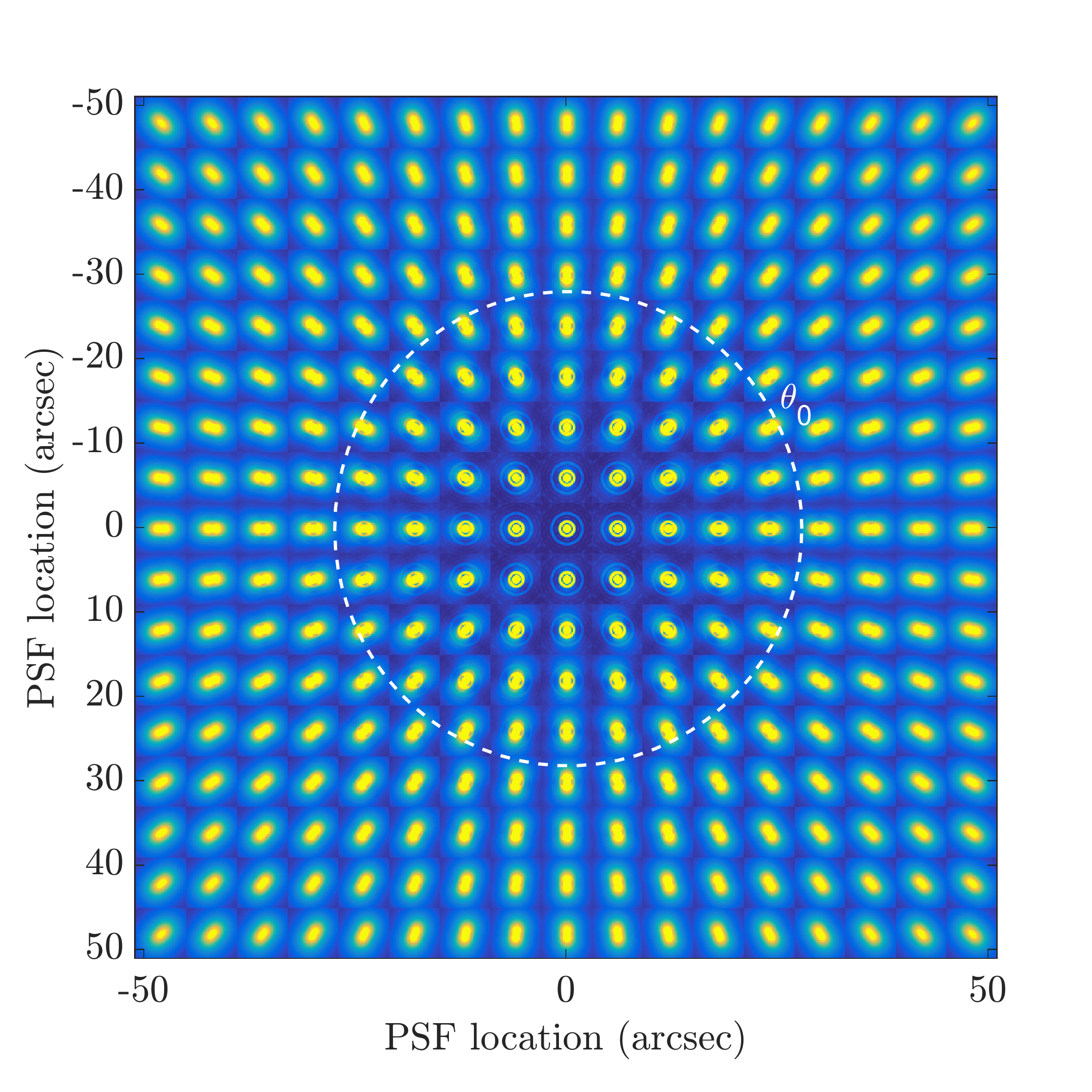}	
		\includegraphics[width=8.8cm]{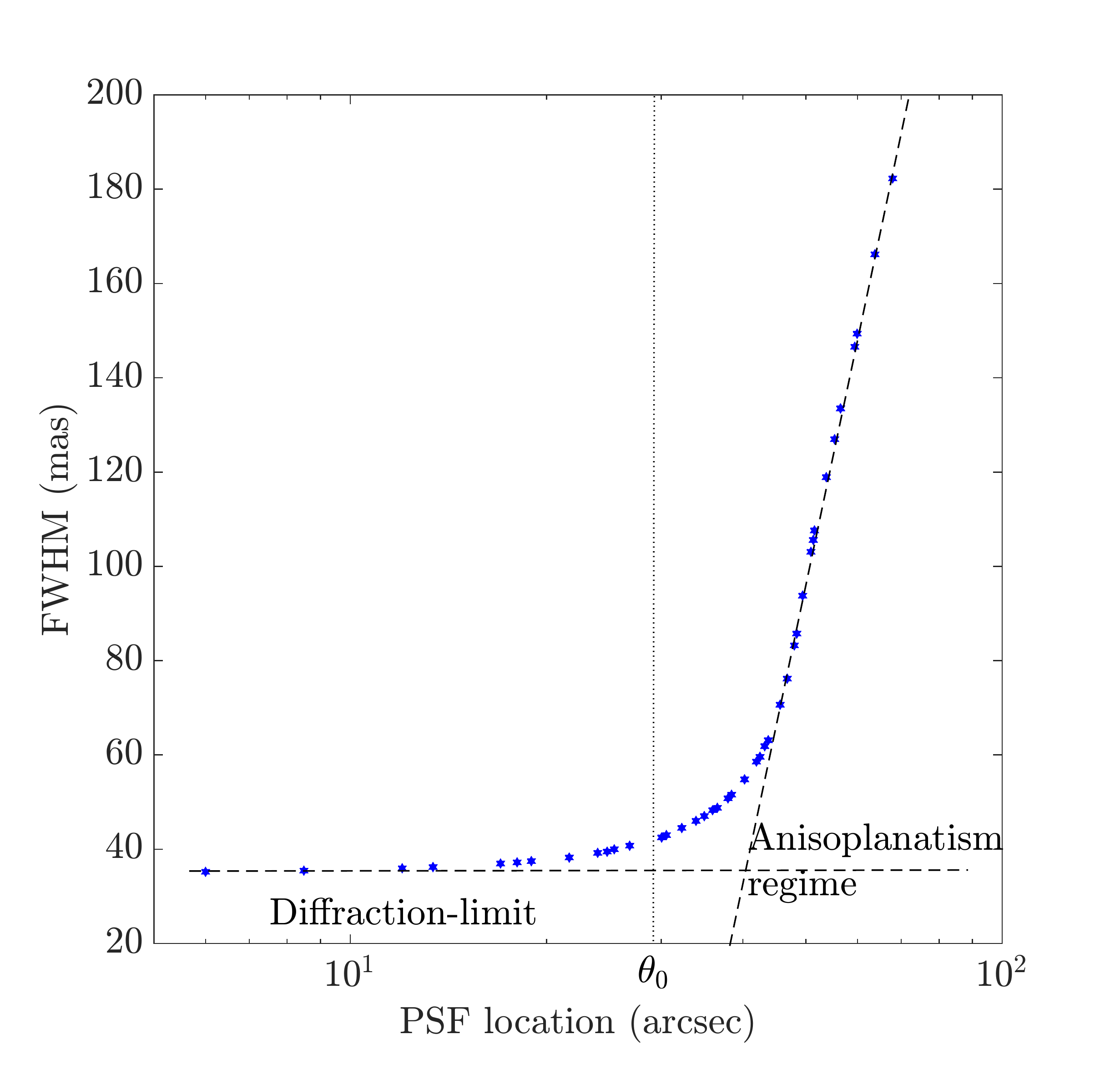}
		\caption{\textbf{Left~:} AO-compensated PSFs across the field. Ticks are giving the corresponding PSF location from the guide star direction. \textbf{Right~:} FWHM as function of field position from on-axis.}
		\label{F:PSFgrid}
	\end{figure}

	As illustrated by Eq.~\ref{E:criteria}, FPP takes benefits from pixels intensity variations in the focal plane and more particularly on a feature that is spatially correlated. When the PSF is off-axis, a large part of this spatial variation is induced by anisoplanatism as illustrated in Fig.~\ref{F:PSFgrid}.
	 \textcolor{black}{We could potentially rely on the FWHM value only to estimate the original $\cnh$ that explains the PSF broadening, but we see several main problems. Firstly, the PSF FWHM varies very little regarding $\theta$ as long as $\theta$ is lower than  $1.5\times \theta_0$. It would consequently require a large field, basically 30" in H-band, to have a linear dependency between the FWHM and the $\cnh$ value, which can be larger than the imager FOV. On top of that, the FWHM is only a scalar value and we need to retrieve the $\cnh$ value over several bins, at least seven as mentioned in \cite{BeltramoMartin2018}; as a consequence we would need at least seven PSF in different position in the field to expect a full profile retrieval, which can be quite challenging. The use of all the PSF pixels within the AO control radius ensure to have theoretically enough sensibility to extract the profile from a single PSF.}\\

	\section{Method sensitivity and performance}
	\label{S:simu}
	
	We have simulated a grid of PSFs presented in Fig.~\ref{F:PSFgrid} using the wavefront propagation object-oriented code OOMAO~\cite{Conan2014OOMAO}. Atmosphere set up refers to Mauna Kea median conditions, telescope and AO characteristics follows a Keck II baseline as detailed in Tab.~\ref{T:setup}. 
	\textcolor{black}{We did not have simulated a real AO system, we have assumed that the on-axis PSF is diffraction-limited and defined by a circular pupil only. We want to figure out the potential of the FPP method in such a limit case before applying it to real AO sky data. More generally, FPP does not need any assumption about the AO system thanks to the knowledge of $\otf{0}$ that contains exactly what the AO system has done. It only cares about the atmospherical turbulence distribution that makes the PSF varies across the field. In other words, simulating a real on-axis PSF would not have really change the results we are showing below, because we assume to know $\otf{0}$ perfectly. The case we do not do this assumption anymore and use PSF reconstruction to identify $\otf{0}$ is treated is another publication.}\\
	
	\begin{table}
		\centering
		\caption{Simulation set up summary based on Mauna Kea median profile. Note that the anisoplanatism accounts on layer bin strictly above 0 km only. }
		\begin{tabular}{l|l}
			Sources wavelength [$\mu$m] & 1.65  \\
			Exposure time [s] & 30  \\
			Zero point [mag/s] & 25.5\\
			$r_0$ [nm] (500nm)      & 0.16  \\       
			$L_0$ [m]               & 25  \\
			$\theta_0$ (1.65$\mu$m) & 22"\\                    
			fractional $r_0$ [\%] 	& 51.7, 11.9, 6.3, 6.1, 10.5, 8.1, 5.4 \\
			altitude layer [km]     & 0, 0.5, 1, 2, 4, 8, 16\\
			Telescope diameter [m]   	& 10\\
			Telescope elevation [$\deg$]  & 30\\
			\# pixels in the pupil  & 200\\
			DM actuator pitch [m]     	& 0.5\\
			WFS \# lenslets       	& 20\\	
		\end{tabular}	
		\label{T:setup}
	\end{table}
	
\begin{figure}
    \centering
    \includegraphics[width = 9cm]{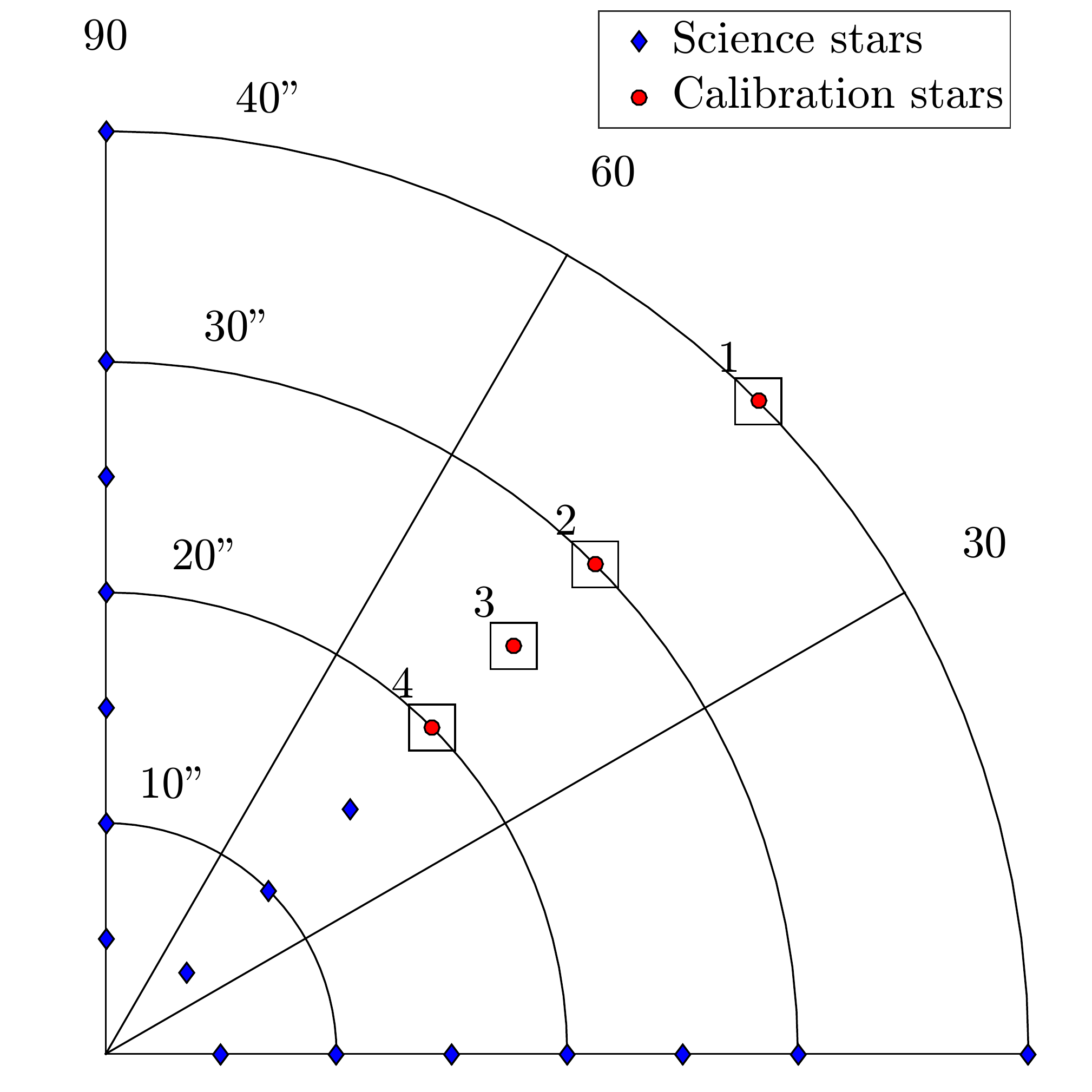}
    \caption{Spatial distribution of simulated sources in the (azimuth angle, zenith angle) plan. Circle markers refer to calibration stars position where the FPP picks off PSFs, in an ordered range from labelled 1 up to 4, to calibrate the $\cnh$ profile.}
    \label{F:polar}
    \end{figure}

    In the following, we compare the FPP $\cnh$ retrieval regarding the calibration star magnitudes and numbers. \textcolor{black}{In all the following, the term calibration will refer to the $\cnh$ estimation using FPP that performs this calibration. For instance, \emph{calibration stars} points the specific stars in the field whose the PSF is provided to FPP to identify the $\cnh$.}    
     Simulated PSFs are distributed along concentric rings of 5" up to 40"-Zenith Angle (ZA) as represented in Fig.~\ref{F:polar}. 
    \textcolor{black}{In Sect.~\ref{SS:sensitivity}, we asses the limit on the calibration field position to ensure a full $\cnh$ retrieval. In Sect.~\ref{SS:noise}, we perform an analyses about how the noise that contaminates calibration PSFs does propagate into the $\cnh$ retrieval process. To do so, we have picked-off calibration PSFs from 40" to 20"-ZA to feed the FPP. To increase the number of non-redundant information and the overall Signal to Noise Ratio (SNR), we have concatenated these PSF from the farthest to the closest, i.e. in Eq.~\ref{E:criteria}, $i=1$ refers to the 40"-ZA star. As long as the calibration PSF is located beyond $\theta_0$, we have verified that the FPP can get back to the entire profile using this PSF, i.e. there is no specific gain by relying on the sole 20"-ZA or the 40"-ZA star, at least for noise-free images. When introducing noise, the 40"-ZA star is more elongated and its energy is spread over the focal plane and it is so more contaminated by the Poisson noise. By getting from the 40"-ZA to 20"-ZA star, we increase progressively the overall SNR.}\\
    
    Stars magnitudes are set up by scaling the PSF flux with regard to the H-band zeropoint value  of 25.5 mag/s for the NIRC2 detector at Keck II. We consider 38 counts by pixel for the read-out noise and 0.08 count/pixel/sec for the dark current. Photon noise is included as well as the sky background as well that reaches 13.6 mag/arcsec$^2$/s in H-band.\\
    
    \textcolor{black}{Finally, in all the following, both simulation and FPP model have the same altitude resolution capability, i.e. we configured the FPP to retrieve the 6 layers in altitude. The impact of altitude distribution errors is discussed in \cite{BeltramoMartin2018}. Because we assume to know $\otf{0}$, the ground layer fraction at 0 km does not participate to the anisoplanatism and can not be identified by FPP. We will present in an future publication an extension to FPP in order to estimate the ground layer contribution as well by adjusting the PSF wings.}

   \subsection{Sensitivity to anisoplanatism strength}
	\label{SS:sensitivity}
	
	We consider that atmosphere bins heights are perfectly known. We firstly address the question of FPP sensitivity: how large must be the field of view to be sufficiently affected by anisoplanatism? \textcolor{black}{Closest PSFs are less contaminated by anisoplanatism as illustrated in Fig.~\ref{F:PSFgrid}, i.e. there are less pixels the FPP can rely on to perform the retrieval. We thus expect that it exists a threshold on $\theta$ below which the FPP can not identify accurately the entire profile.}\\
	
\begin{figure}
		\centering
		\includegraphics[width=16cm]{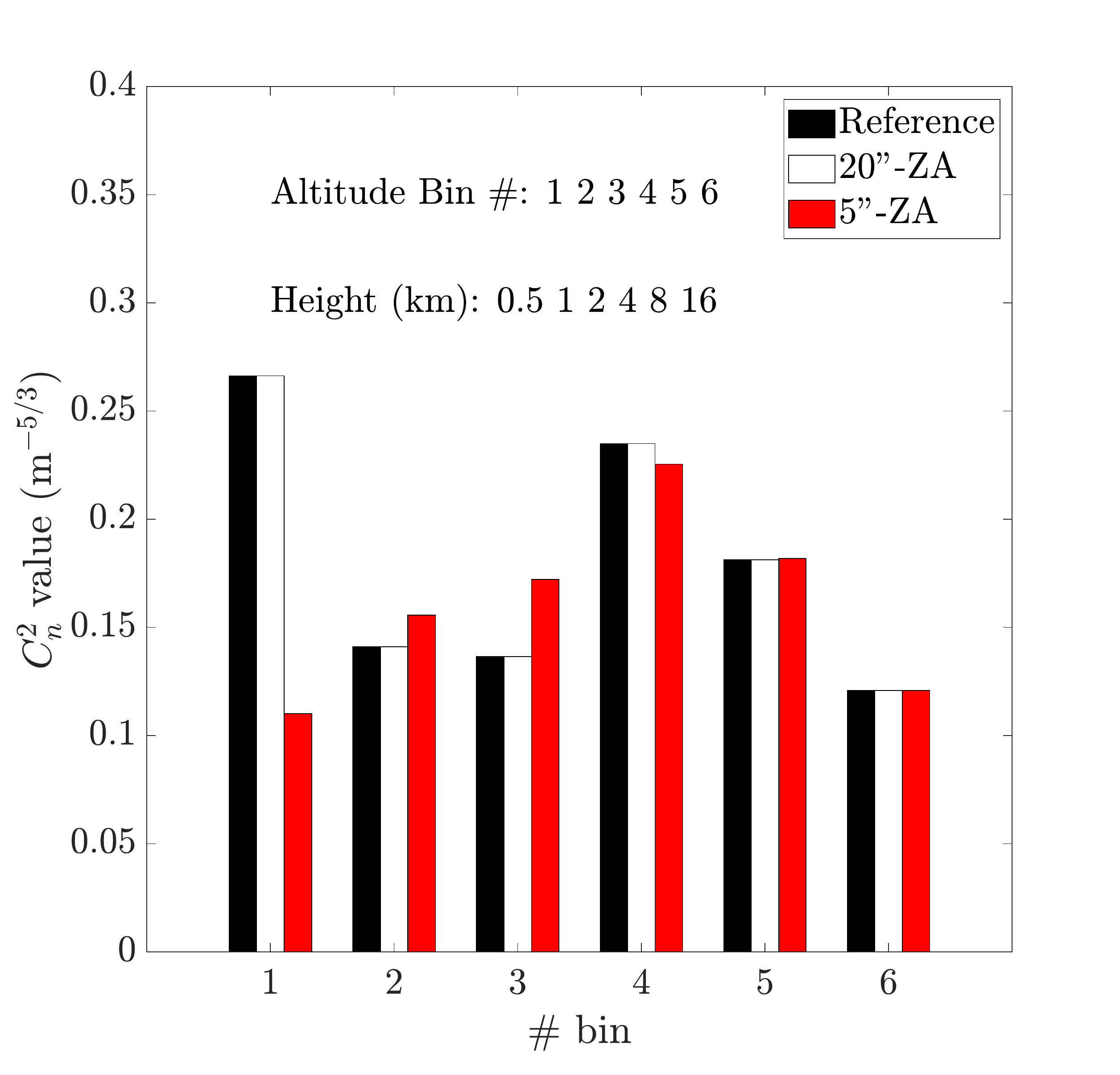}
		\caption{Retrieved $\cnh$ profile using a single PSF located either at 5" or 20"-ZA.}
		\label{F:cnhVloc}
	\end{figure}

	To figure out the existence of this threshold, we have run the FPP algorithm to retrieve the $\cnh$ using a single simulated PSF located between 1" up to 40" from the GS position. We report in Fig.~\ref{F:cnhVloc} a comparison of estimated $\cnh$ regarding the PSF location. It shows that the profile is perfectly identified as long as the PSF is sufficiently affected by the anisoplanatism. \\
    
\begin{figure}
   \centering
   \includegraphics[width=16cm]{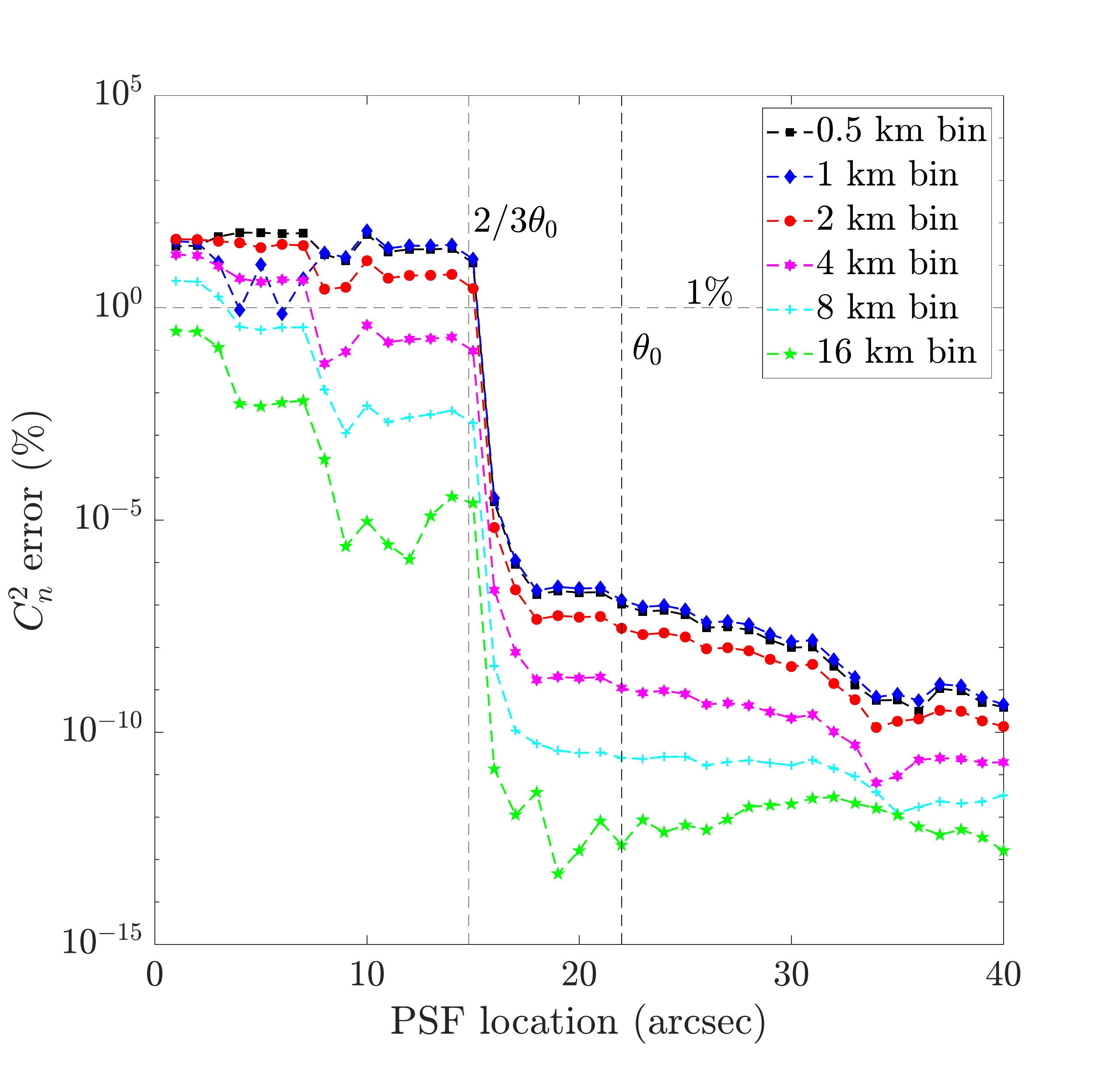}
   \caption{Accuracy on $\cnh$ bins regarding the calibration PSF location that is provided to the FPP.}
   \label{F:cnhVlocnonoise}
   \end{figure}

    Its is confirmed by Fig.~\ref{F:cnhVlocnonoise} that shows the accuracy on bins strengths regarding the calibration star position. It does spectacularly highlight that the entire profile is retrieved as long as $\theta > 2/3\theta_0$. 
    Below this threshold, bins at the lowest altitude are not well retrieved, with a clear trend of accuracy degradation for shorter PSF location and lower altitude. 
\textcolor{black}{We notice that the accuracy improvement is not linear with respect to $\theta$, but resembles more like to a succession of plateaus. We believe is due to the particular Mauna Kea median profile that we have used to perform this study; if we took another one we would have a different shape, but still the same level of mean error and a strong breakout at $\theta = 2/3\theta_0$, in a way it does not change the main conclusion of our analyses.} \\   
    
    Fig.~\ref{F:cnhVlocnonoise} gives also the number of layers we can reconstruct with regard to the location we pick-off the PSF. For instance, to get 1\% of accuracy on the $\cnh$ using a 10"-ZA PSF, we can only reconstruct the three highest layers, down to two for a 5"-ZA PSF.
    In other words, in the case of a 10"-FOV, as for the narrow-field mode of NIRC2, a 3 altitude layers-based PSF model is sufficient to characterize its properties at 1\%-level accuracy. 
 
	We report in Tab.~\ref{T:th0} the isoplanatic angle value as function of the $\rz$ value at zenith and the telescope elevation for an outer scale of 25~m. Considering we need a PSF positioned $\theta_0$ off-axis, it informs about where we must pick-off an observation to ensure the full $\cnh$ retrieval at 1\% accuracy. For instance, for median conditions ($\rz$ = 16 cm), we would need an observation at 22" to retrieve 7 layers for 30 degree of telescope elevation.
	
	\begin{table}
		\centering	
		\caption{Isoplanatic angle values at 1.65 $\mu$m regarding the $r_0$ zenith value at 500 nm at Mauna Kea and the telescope elevation. The outer scale was fixed to 25m.}
		\begin{tabular}{|c|c|c|c|c|c|}
			
			\hline
			& \multicolumn{5}{c|}{Telescope elevation [deg]}\\
			\hline
			$\rz@500$nm [cm]    & 0 & 15 & 30 & 45 & 60 \\	
			\hline
			8 & 12.1& 11.4 & 9.5 & 6.7 & 3.8 \\
			\hline
			12 & 19.8& 18.6& 15.0 & 10.9 & 6.0\\
			\hline
			16 & 28.6& 26.9& 22.0 &15.5  & 8.5\\
			\hline
			18 & 33.5& 31.4& 26.1 & 18.0& 9.8\\
			\hline	
			20 & 38.7& 36.3& 30.2 & 20.6 & 11.2\\
			\hline	
		\end{tabular}
		\label{T:th0}
	\end{table}
				
	\subsection{Sensitivity to noise}
	\label{SS:noise}
	
	We now consider a 40"-field case~(NIRC2 wide-field mode), with the atmosphere/telescope set up displayed in Tab.~\ref{T:setup}. We address the point about how noise propagation affects the $\cnh$ retrieval. 
	
	We keep handling simulated images, that are now contaminated following a NIRC2 baseline. As we saw previously, these PSFs are anisoplanatism-contaminated enough to permit a full profile retrieval over the 7 layers. \\
	
\begin{figure}
		\centering
		\includegraphics[width=16cm]{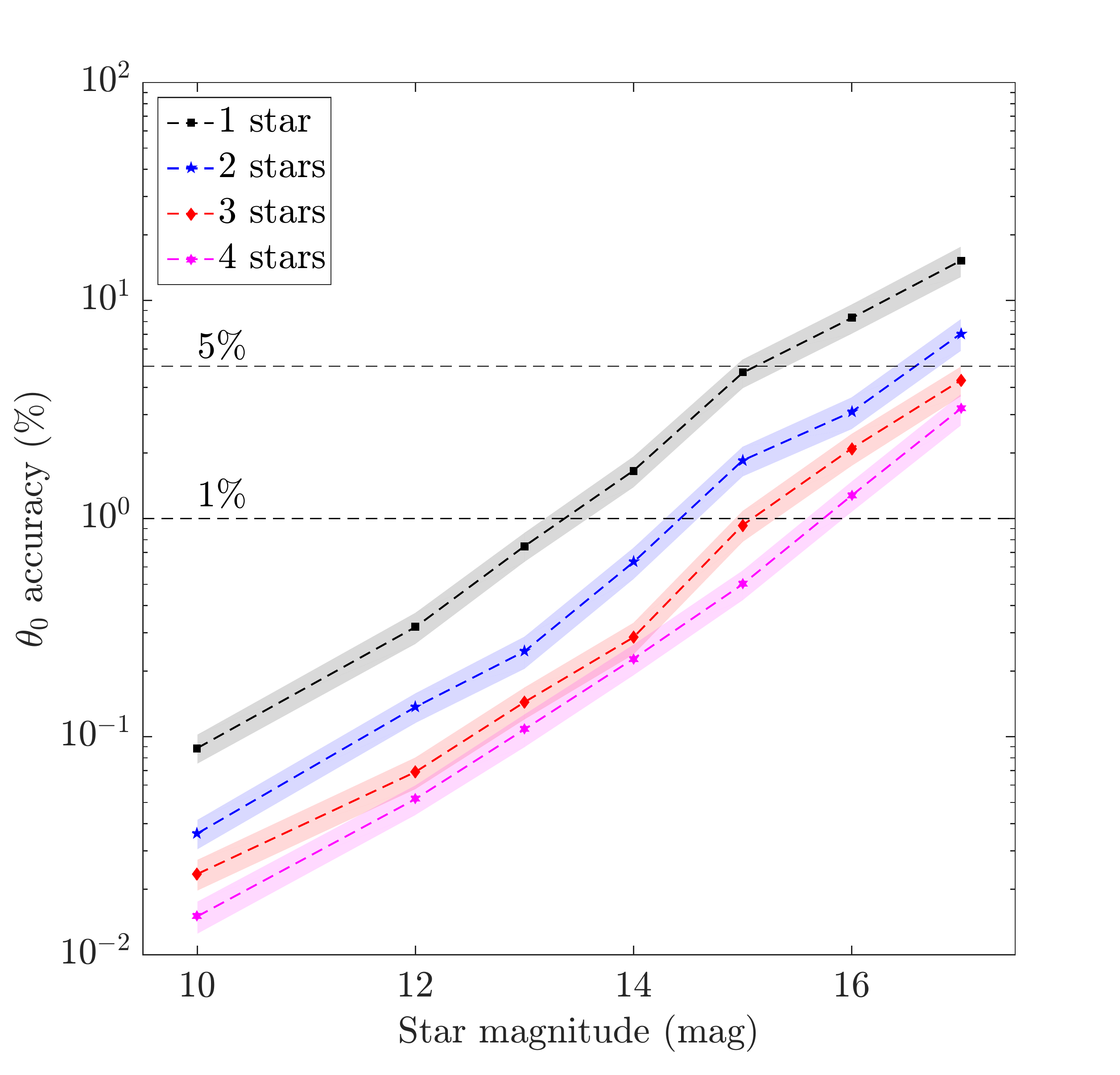}
		\caption{H-band isoplanatic angle $\theta_0$ accuracy as function of the FPP calibration stars magnitudes. Envelopes are given from a 1-$\sigma$ standard-deviation.  }
		\label{F:th0Vnpsf}
	\end{figure}

	The FPP algorithm is tested for a collection of PSFs whose magnitude~(the same for each star) varied from H=10 to 17. We report in Fig.~\ref{F:th0Vnpsf} the FPP-estimated $\theta_0$ as function of the calibration stars magnitudes and numbers. It illustrates that the noise that contaminates calibration PSF propagates through the minimisation process and does degrade the $\cnh$ retrieval. This contamination can be fought and mitigated by relying on more calibration PSFs, which improves the overall SNR and the $\cnh$ accuracy. 1\% of accuracy on $\theta_0$ is roughly obtained with a single star of magnitude 13.5.

\begin{figure}
		\centering
		\includegraphics[width=16cm]{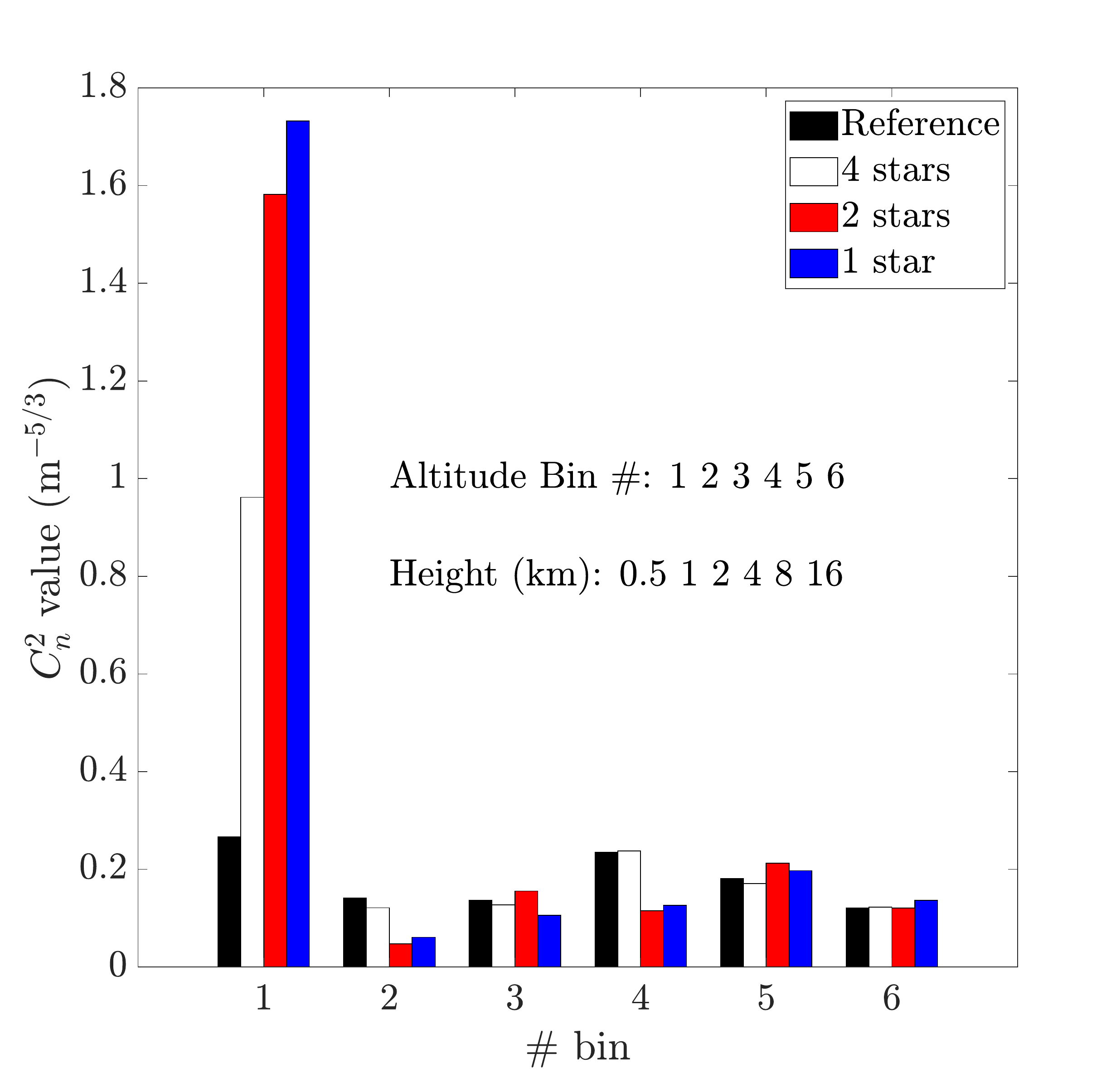}
		\caption{Retrieved $\cnh$ profile as function of the number of collected PSFs for calibration stars magnitudes set to 15.}
		\label{F:cn2hVnpsf}
	\end{figure}

\begin{figure}
		\centering
		\includegraphics[width=16cm]{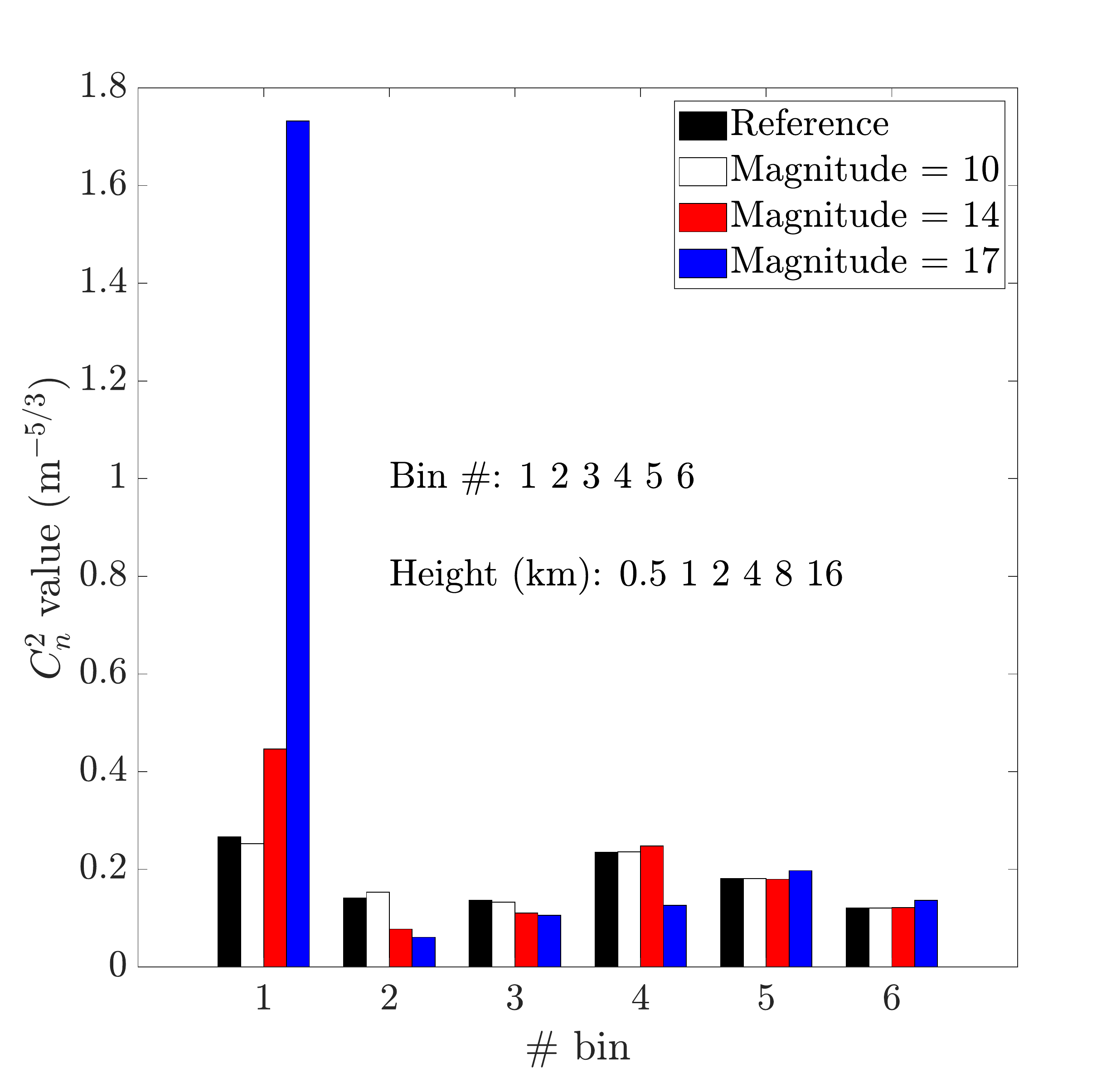}
		\caption{Retrieved $\cnh$ profile with respect to calibration stars magnitudes when picking off 4 PSFs in the field.}
		\label{F:cn2hVmag}
	\end{figure}

	We report in Figs.~\ref{F:cn2hVnpsf} and~\ref{F:cn2hVmag} the FPP-estimated $\cnh$ profile compared to the simulation reference for different numbers of collected PSFs and different magnitude levels. As previously, results confirm that increasing the number of calibration stars does help the identification. Furthermore, the lowest bins are the most sensitive to the noise level; they are the ones that contribute the least to anisoplanatism effect as discussed in Sect.~\ref{SS:sensitivity}. The degradation of $\theta_0$ is consequently mostly explained by inaccuracies on these lowest layers. 
    
    As a conclusion, in order to provide a $\cnh$ profile estimation that reaches 1\% of accuracy on $\theta_0$, the FPP needs to rely on a single-star of magnitude H=13.5, located at least beyond $2/3\times\theta_0$ and regarding the NIRC2 baseline we have considered in this study. We expand this quantitative constrain of the FPP as function of the exposition time and number of calibration stars in Sect.~\ref{SS:threshold}.


	\section{Application to PSF extrapolation}
	
	We still refer to our simulation baseline in order to assess how accurate the FPP permits the PSF extrapolation. As previously, we picked-off several stars in a anisoplanatism-limited area, with noise contamination, to calibrate the PSF model across all the field. Modelled PSFs are compared to simulation in terms of Strehl-ratio~(SR) and Full width at half maximum~(FWHM).\\
	
\begin{figure*}
	\centering
    \includegraphics[width=18cm]{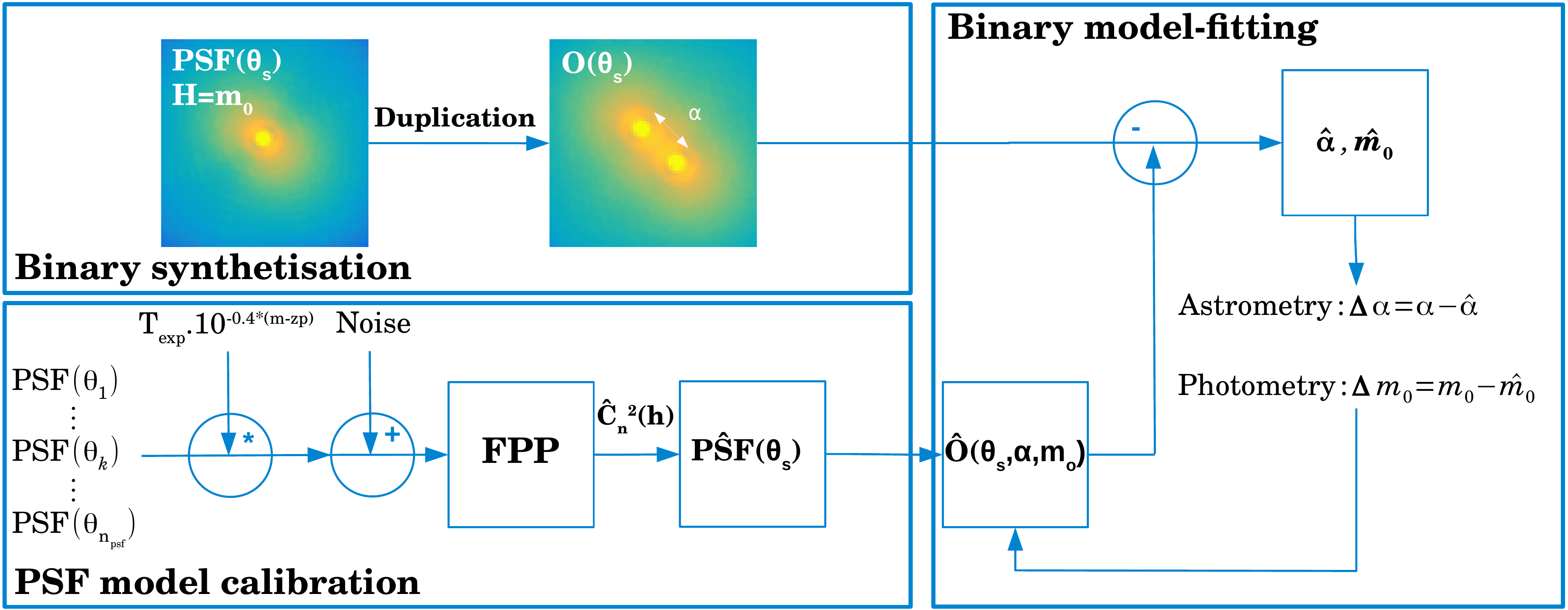}
    \caption{Sketch of the binary parameters retrieval. For each field location $\theta_s$, we create a fake binary with a known separation $\alpha$ and magnitude $m_0$. In parallel, we calibrate the PSF model using the FPP over a collection of noise-contaminated calibration PSFs of magnitude $m$, with zp the zero-point and T$_\text{exp}$ the exposure time. This model is duplicated to create a binary-model as function of the stars fluxes and relative separation, which is adjusted iteratively over the synthetic observation. }
    \label{F:bin}
	\end{figure*}

	On top of that, we aim at quantifying astronomical-related metrics, such as photometry and astrometry. We followed the same baseline as presented in~\cite{BeltramoMartin2018} for tight binaries. 
    For each simulated star, we create a corresponding 100 mas-separated fake binaries by duplicating and shifting the PSF. The purpose of this manipulation is to measure how well we can retrieve the binaries characteristics by fitting a model provided by the FPP. \textcolor{black}{Basically, photometry and astrometry are measured from the residual of the PSFs scaling and relative positions adjustments over the synthetic binary, when using a binary model based on the FPP PSF model that may differ from the real simulated PSF. The process is repeated for any value of $\theta$.}
    Because the FPP relies on noisy calibration PSFs, we aim here at assessing how the noise impacts the binaries parameters estimation by biasing the PSF-model representation that we calibrate using the FPP. We summarise the methodology in Fig.~\ref{F:bin}.\\
            
	For each considered metric, we look either at its mean value across the field regarding the calibration stars magnitudes, or how it does vary with angular separation from on-axis.
	
	\subsection{Impact on PSF morphology}
	\label{SS:sr}
	
    SR and FWHM are known~\cite{Roddier1999} to follow $\theta_0$-dependent laws given respectively by $\exp(\theta_0^{-5/3})$ and $\para{\theta/\theta_0}^{-5/3}$. We expect to retrieve trends connected to what is presented in Fig.~\ref{F:th0Vnpsf}.\\

\begin{figure}
		\centering
		\includegraphics[width=16cm]{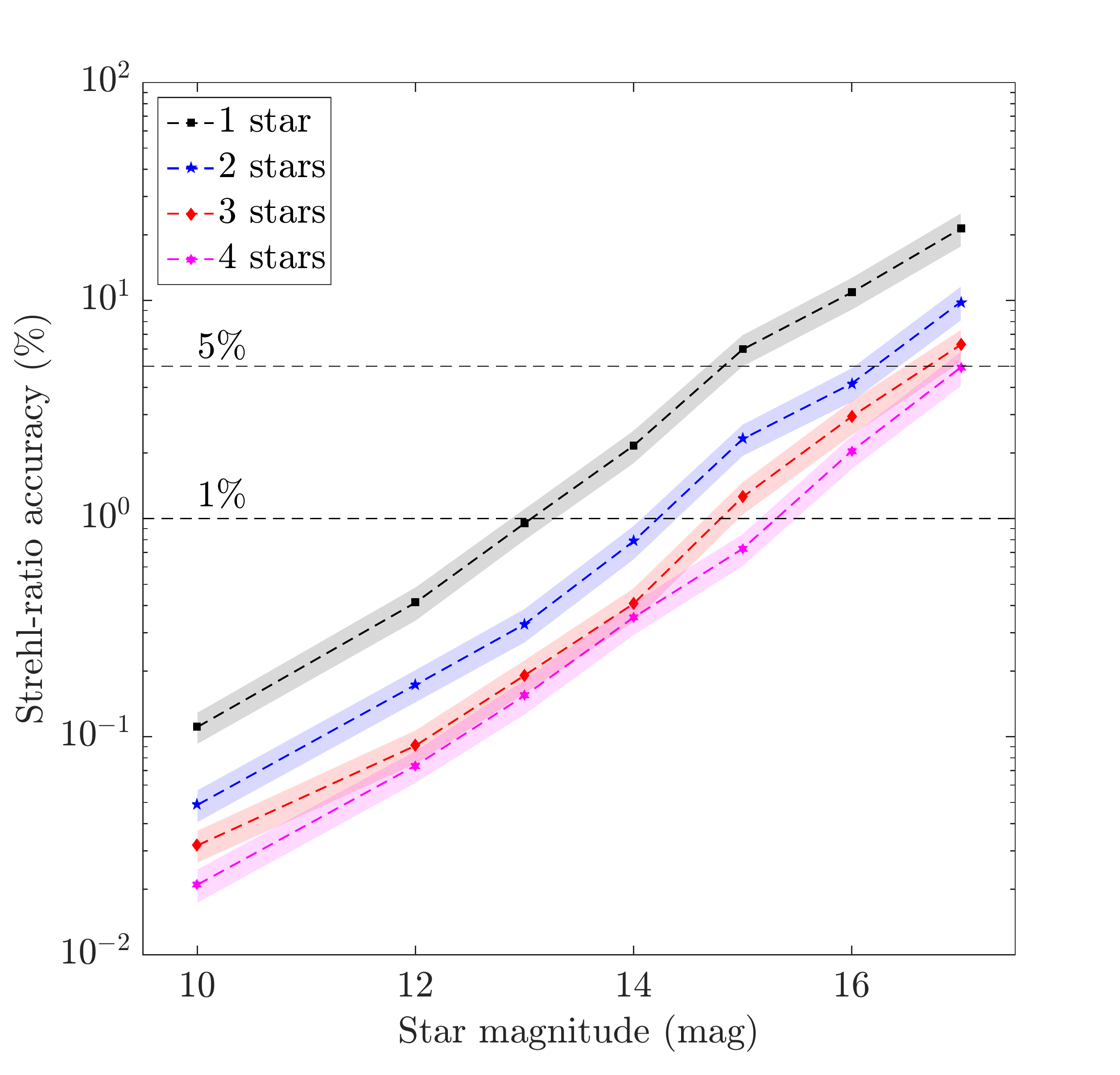}
		\caption{H-band SR accuracy averaged out over the field regarding the calibration stars magnitudes. Envelopes are given from a 1-$\sigma$ standard-deviation.}
		\label{F:srVmag}
	\end{figure}

\begin{figure}
		\centering
		\includegraphics[width=16cm]{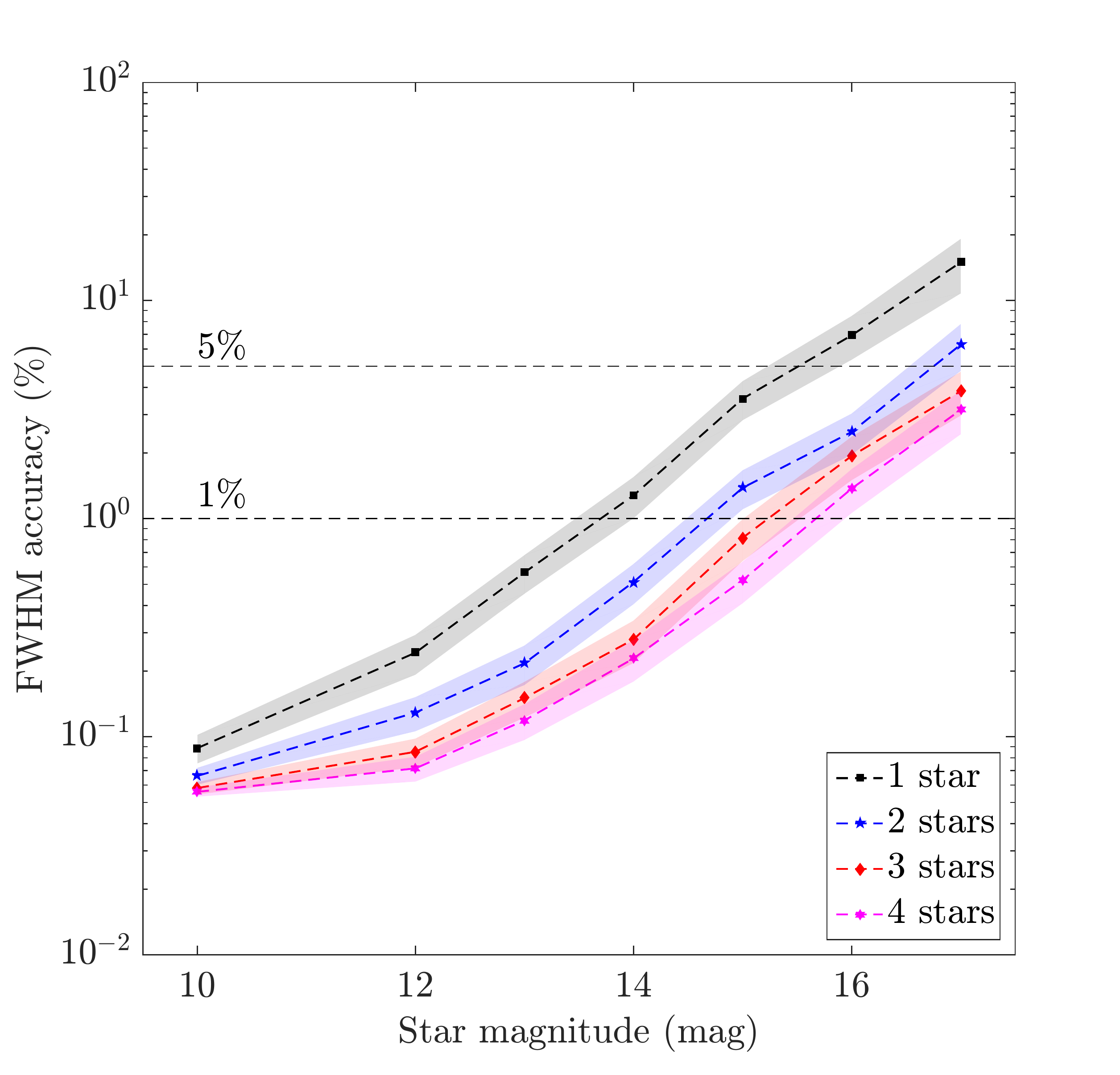}
		\caption{H-band FWHM accuracy averaged out over the field regarding the calibration stars magnitudes. Envelopes are given from a 1-$\sigma$ standard-deviation.}
		\label{F:fwhmVmag}
	\end{figure}

    Figs.~\ref{F:srVmag},~\ref{F:fwhmVmag} display SR and FWHM mean accuracy in the field with regard to the calibration stars magnitudes. We retrieve the similar linear trends with respect to $\theta_0$ as presented in Fig.~\ref{F:th0Vnpsf}, that confirms that accuracy of those parameters are connected to $\theta_0$ estimation errors regarding the noise level.\\
	

\begin{figure}
		\centering
		\includegraphics[width=16cm]{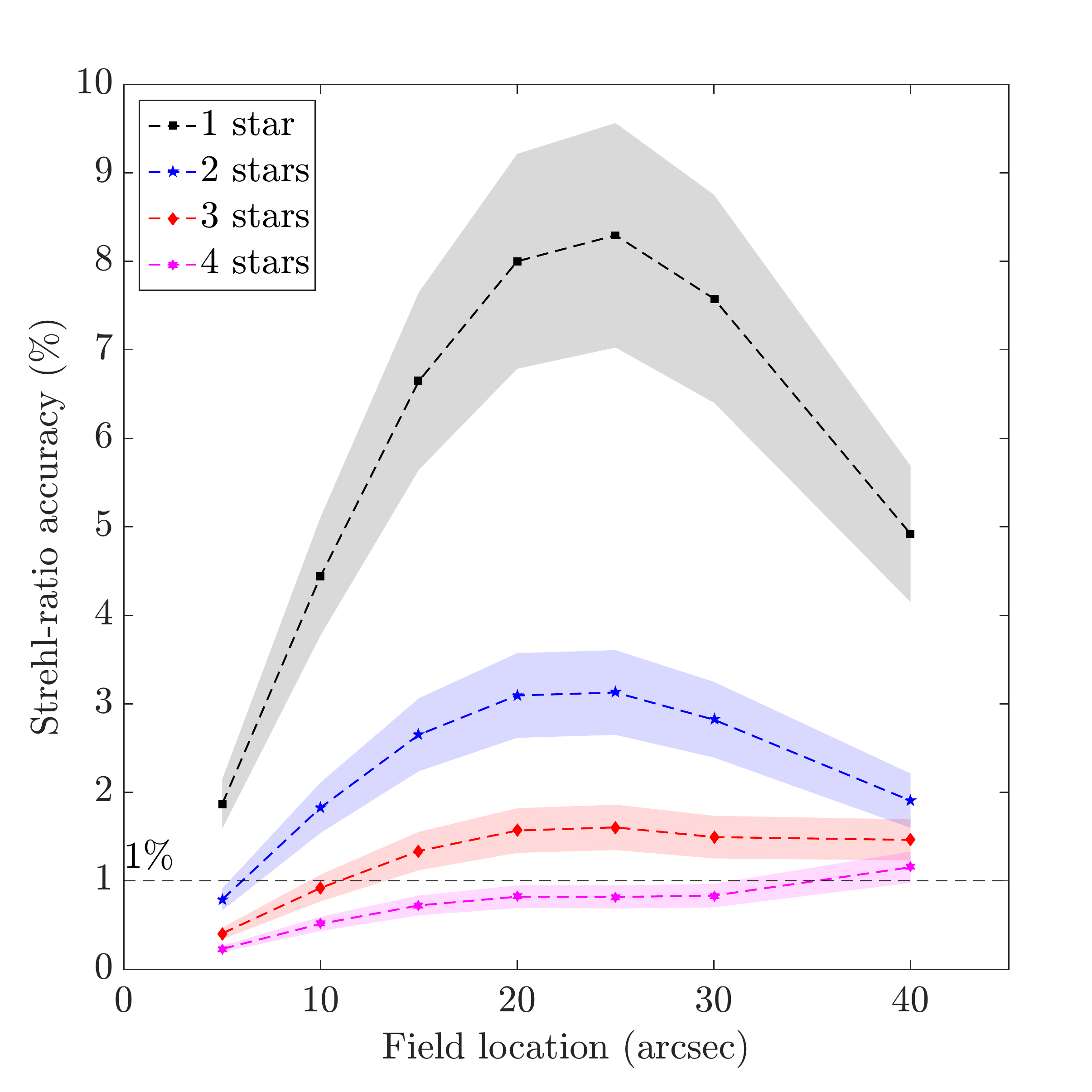}
		\caption{H-band SR accuracy in the field regarding number of calibration stars of magnitude 15. Envelopes are given from a 1-$\sigma$ standard-deviation.}
		\label{F:srVnpsf}
	\end{figure}

\begin{figure}
		\centering
		\includegraphics[width=16cm]{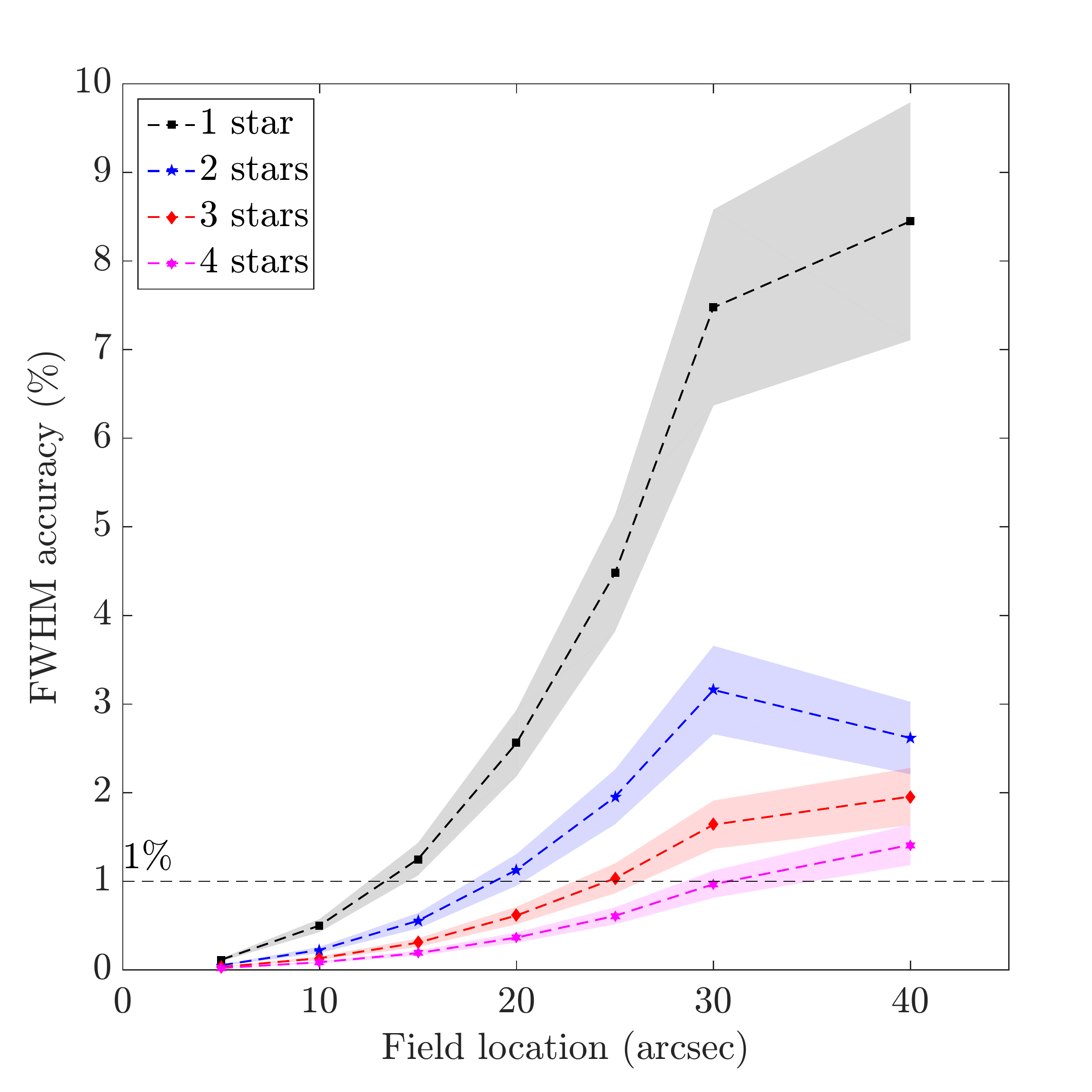}
		\caption{H-band FWHM accuracy in the field regarding number of calibration stars of magnitude 15. Envelopes are given from a 1-$\sigma$ standard-deviation.}
		\label{F:fwhmVnpsf}
	\end{figure}

	We illustrate in Figs.~\ref{F:srVnpsf} and~\ref{F:fwhmVnpsf}, SR and FWHM accuracy as function of $\theta$ when calibrating the $\cnh$ profile with 1 up to 4 stars of magnitude 15.  When relying on more stars, the $\cnh$ estimation is better as seen in Sect.~\ref{SS:noise}, which translates into a better PSF modelling downstream.\\
	
	Nevertheless, the model accuracy is not uniform across the field. Closest stars images from on-axis are the less affected by anisoplanatism, i.e. the PSF at those position are not sensitive to any $\cnh$ mis-retrieval, which explains why SR and FWHM values get better estimated for weaker separation. Then, the PSF model is calibrated using at least the 40"-ZA star; consequently the PSF must be well characterised at this specific location for the single calibration PSF case, which justifies why we see this drop in error for the 40" separation. 
    By gathering up more calibration stars from 40"-ZA to 20"-ZA, in addition to improve the PSF morphology characterisation, we also make estimates more uniform across the field. 

	\subsection{Impact on binaries photometry and astrometry}			
	
\begin{figure}
		\centering
		\includegraphics[width=16cm]{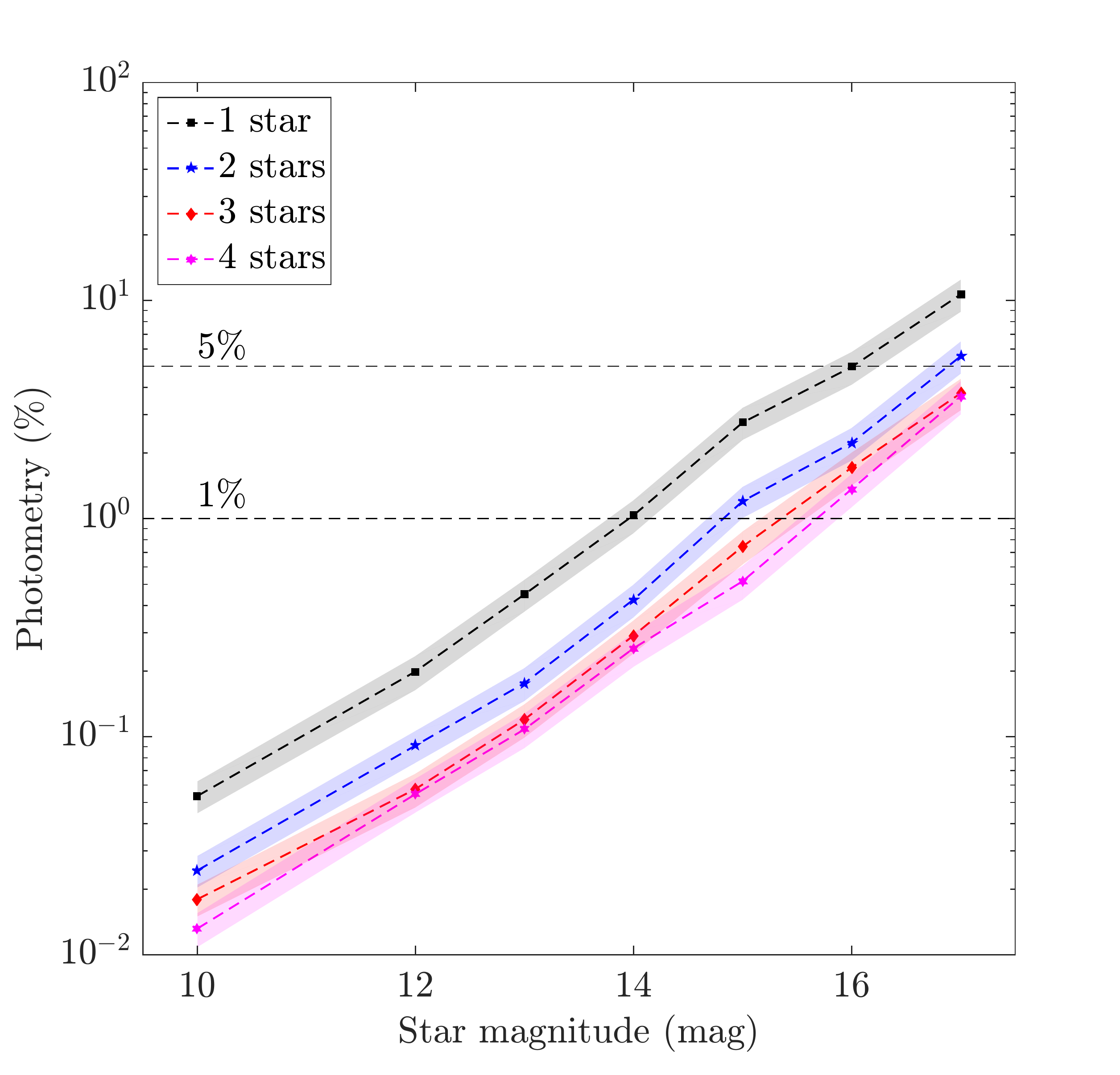}
		\caption{H-band photometry accuracy averaged out over the field regarding the calibration stars magnitudes. Envelopes are given from 1-$\sigma$ standard-deviation.}
		\label{F:photoVmag}
	\end{figure}

\begin{figure}
		\centering
		\includegraphics[width=16cm]{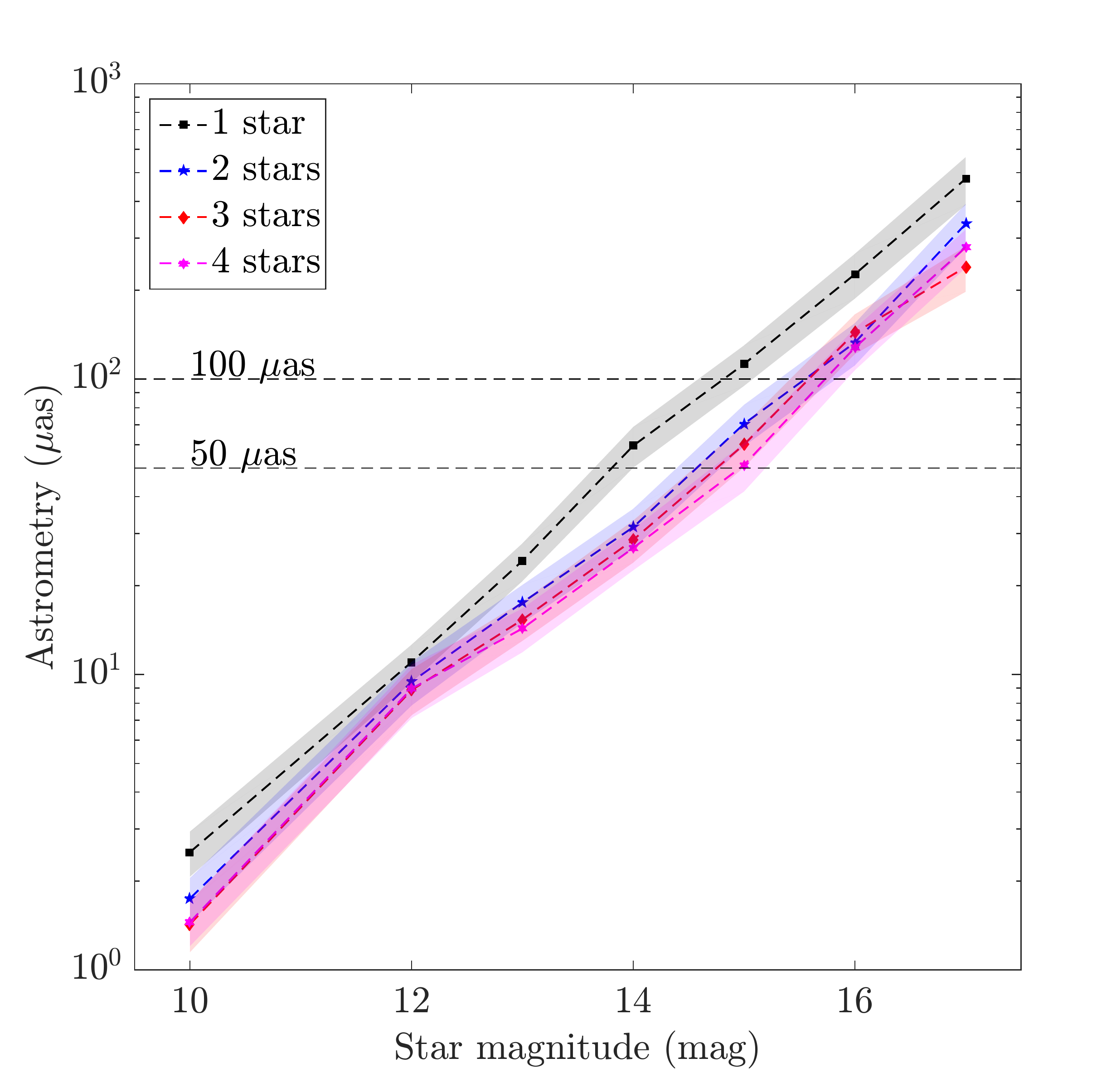}
		\caption{H-band astrometry accuracy averaged out over the field regarding the calibration stars magnitudes. Envelopes are given from 1-$\sigma$ standard-deviation.}
		\label{F:astroVmag}
	\end{figure}

\begin{figure}
		\centering
		\includegraphics[width=16cm]{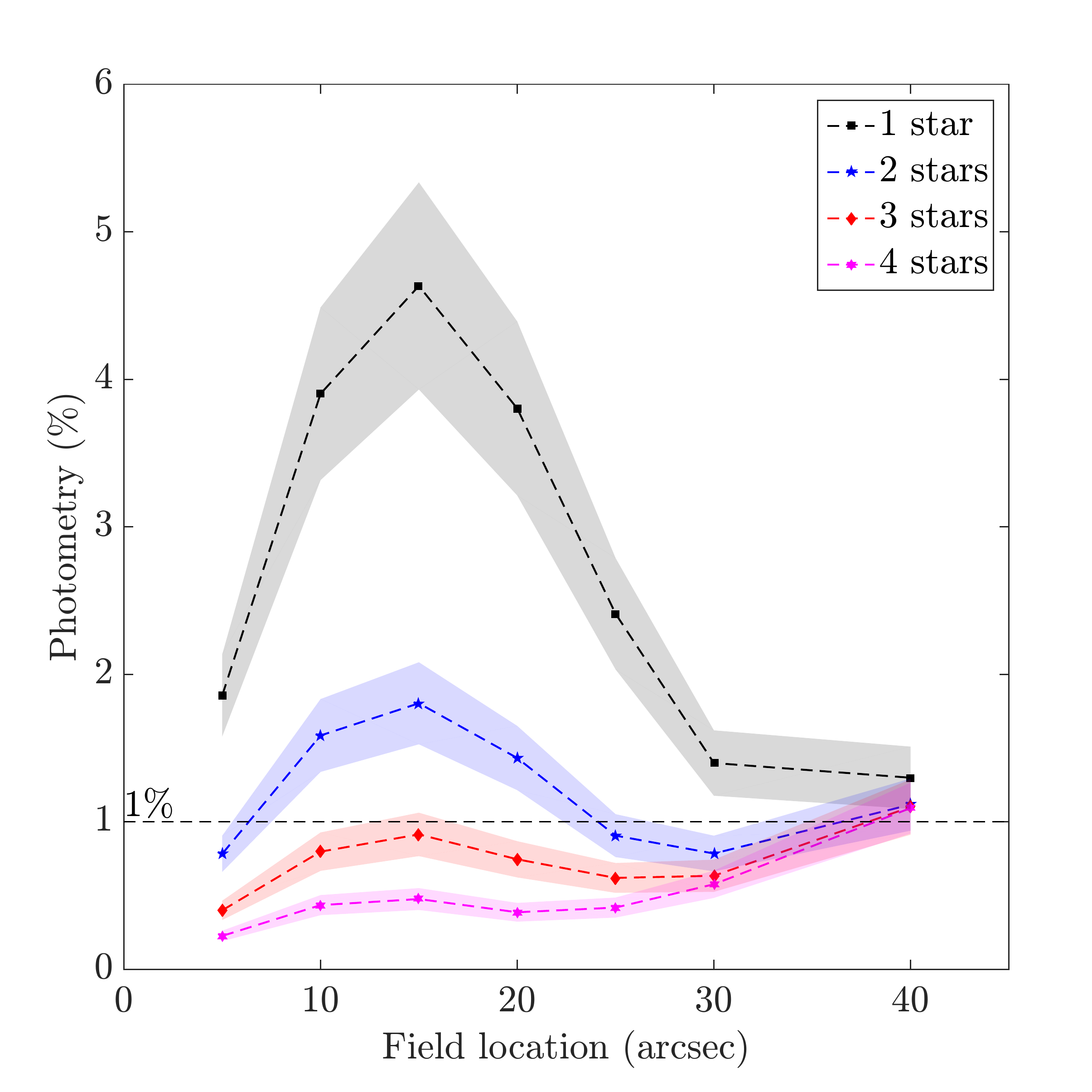}
		\caption{H-band photometry accuracy in the field regarding number of calibration stars of magnitude 15. Envelopes are given from 1-$\sigma$ standard-deviation.}
		\label{F:photoVtheta}
	\end{figure}

\begin{figure}
		\centering
		\includegraphics[width=16cm]{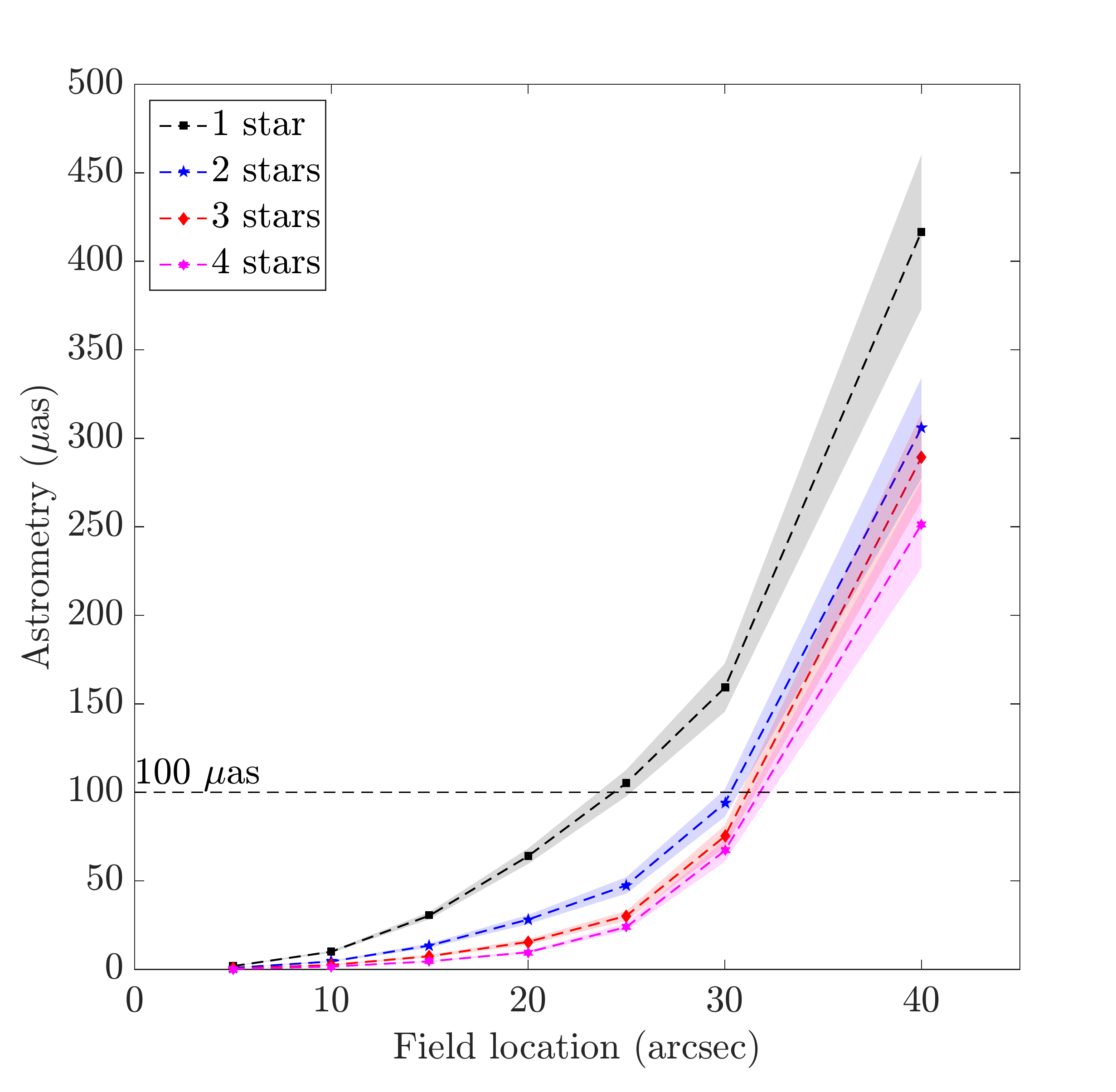}
		\caption{H-band astrometry accuracy in the field regarding number of calibration stars of magnitude 15. Envelopes are given from 1-$\sigma$ standard-deviation.}
		\label{F:astroVtheta}
	\end{figure}

	We report in Figs.~\ref{F:photoVtheta} and~\ref{F:astroVtheta} respectively the photometry and astrometry with respect to the PSF location in the field. As for PSF-related metrics, photometry and astrometry are not estimated uniformly for the same reasons as explained in Sect.~\ref{SS:sr}. Figs.~\ref{F:photoVmag} and~\ref{F:astroVmag}  show the same linear trends on science estimates errors with respect to calibration stars magnitudes.\\
	
	As a matter of fact, we observe that photometry and astrometry accuracy are respectively connected to SR and FWHM estimation as we were expecting~\cite{BeltramoMartin2018}. For a single calibration star of magnitude 14, we obtain 1\%-level of photometry and 50$\mu$as of astrometry. These values reflect only PSF-model errors onto the estimates; we do not consider any other potential effects that participate to the overall error breakdown as detailed in~\cite{Fritz2010} for astrometry in the galactic center for instance. It explains why the astrometry drops down to zero for stars close to the guide star: in absence of anisoplanatism, we do not propagate any $\cnh$ mis-retrieval into the PSF model.
    
    Contrary to photometry, astrometry behaves as a monotonic function of the field location, though. It points out that the astrometry is roughly given by the ratio FWHM/SNR and so sensitive to the \textcolor{black}{slope of the averaged PSF core.} Because the PSF FWHM increases monotonically as function of $\theta$ as shown in Fig.~\ref{F:PSFgrid}, we retrieve the same behaviour for the astrometry versus $\theta$, where the function goes up less rapidly if we increase the SNR by calibrating the PSF model over more stars.\\

    One may argue that more optimal tools exist to estimate photometry and astrometry. Yet, our purpose is only to focus on PSF model error contributions due to bad anisoplanatism characterisation, which does not require to deploy such algorithms to be figured out.    
    We highlight these contributions reaches down to $50\mu$as and 1\% respectively of astrometry and photometry over 40" in calibrating the FPP using a single star. In a near-future, we will apply the FPP on real crowded field observations and use standard pipeline to asses the potential gain on estimates.

	
	\subsection{Threshold values}
	\label{SS:threshold}
    
	We benefit from our previous analyses to figure out which configuration permits a 1\%-accuracy estimation of a given metric. We will do the exercise for photometry, but the same methodology is applicable for any other metrics, as long as we have identified how it does vary with regard to the SNR.\\
	
	According to Fig.~\ref{F:photoVmag}, a 1\%-level photometry in the field require to calibrate the FPP model using a single $m_0 = 14$ magnitude PSF imaged with 30~s exposure time. For a different star magnitude, exposure time and number of collected images, we must ensure that the total gathered flux corresponds to a 14 magnitude star. If $m_\star(n_\text{psf},T_\text{exp})$ is the magnitude limit over the $n_\text{psf}$ we get from the field and imaged at $T_\text{exp}$ seconds of exposure time, it must satisfy
	\begin{equation}
		-2.5\times\log_{10}\para{\dfrac{n_\text{psf}\times T_\text{exp}}{30}\times 10^{-0.4\times m_\star}} = m_0
	\end{equation}
	
	that leads to
	\begin{equation} \label{E:maglim}
		m_\star(n_\text{psf},T_\text{exp}) = m_0 + 2.5\times\log_{10}\para{\dfrac{n_\text{psf}\times T_\text{exp}}{30}},
	\end{equation}
	where finally $m_\star$ indicates what must be the maximal magnitude to ensure 1\%-level of photometry estimation across the 40"x40"-field. We report in Tabs.~\ref{T:maglim} results of Eq.~\ref{E:maglim} for couples ($n_\text{psf}$/$T_\text{exp}$).
	We emphasise that these are empirical results obtained for a NIRC2 baseline with a specific electronic noise configuration, but they highlight that deploying the FPP approach is feasible. We will also consider the near infra-red spectro-imager OSIRIS at Keck II that has an 20.4"-FOV imager located at 19" from the spectrograph, which permits to access strongly anisoplanatism-contaminated PSFs.    
	The next step of this work will consist in coupling FPP with on-axis PSF reconstruction to be tested on crowded field images.
	
	\begin{table}
		\centering	
		\caption{H-band stars magnitude limits regarding the number of star and exposure time to get 1\% of accuracy on photometry within 40"$\times$40" FOV using NIRC2 at Keck II. }
		\begin{tabular}{|c|c|c|c|c|c|}
			\hline
			& \multicolumn{5}{c|}{$T_\text{exp}$ [s]}\\
			\hline
			$n_\text{psf}$ & 1 & 10 & 30 & 60 & 120 \\	
			\hline
			1 & 10.3& 12.8 & 14 & 14.7 & 15.5 \\
			\hline
			2 & 11.0& 13.6& 14.7 & 15.5 & 16.3\\
			\hline
			3 & 11.5& 14.0& 15.2 &15.9  & 16.7\\
			\hline
			4 & 11.8& 14.3& 15.5 & 16.3& 17.0\\
			\hline	
			5 & 12.0& 14.6& 15.7 & 16.5 & 17.2\\
			\hline	
		\end{tabular}
		\label{T:maglim}
	\end{table}

	\section{Application to the HeNOS testbed}
	\label{S:henos}
	
	HeNOS (Herzberg NFIRAOS Optical Simulator) is a multi-conjugated AO test bench designed to be a scaled down version of NFIRAOS,
	the first light adaptive optics system for the Thirty Meter Telescope~\cite{Rosensteiner2016}. We used HeNOS in SCAO mode and closed the loop on ones of the 4 LGSs distributed over a 4.5" square constellation, while atmosphere is created using three phase screens. Summary of main parameters is given in Tab.~\ref{T:setupHENOS}. To simulate the expected PSF degradation across the field on NFIRAOS at TMT, all altitudes are stretched up by a factor 11. Moreover, at the time we acquired HeNOS data, science camera was conjugated at the LGSs altitude; LGSs beams are propagating along a cone but are arriving in-focus at the science camera entrance.
	
	\begin{table}
		\centering
		\caption{HENOS set up summary.}
		\begin{tabular}{l|l}
			Asterism side length  & 4.5"\\
			Sources wavelength    & 670 nm \\
			$r_0$ (670 nm)        &   0.751  \\      
			$\theta_0$ (670 nm)	  &  0.854"  \\                   
			fractional $r_0$      &  74.3\% ,17.4\%,8.2\% \\
			altitude layer        & (0.6, 5.2, 16.3) km\\
			source height         & 98.5 km \\
			Telescope diameter    & 8.13~m\\
			DM actuator pitch     & 0.813~m\\
		\end{tabular}	
		\label{T:setupHENOS}
	\end{table}

	We have acquired closed-loop telemetry when guiding the AO system on a LGS in SCAO mode in July 2017. On top of that, we have measured PSFs without phase screens through the beam to characterise best-performance in the current set-up. We report in Tab.~\ref{T:SRHenOS} Strehl-ratio and FWHM measurements. On off-axis PSFs, we see that anisoplanatism degrade significantly the performance, and dominates any other sources of residual errors on off-axis PSFs. This is an ideal situation to test the FPP.
	
	\begin{table}
		\caption{Strehl-ratio and FWHM measured on HeNOS PSFs when closing the loop in a SCAO mode.}
		\centering
		\begin{tabular}{|c|c|c|c|c|}
			\hline
			& PSF on-axis & PSF 1 & PSF 2 & PSF 3 \\
			\hline
			Strehl ratio [\%]& 28.5& 4.0 & 5.1&4.9  \\
			\hline
			No phase screen & 39& 41 & 36& 38  \\
			\hline
			\hline
			FWHM [mas] & 21 & 116& 71 & 100\\		
			\hline
			No phase screen & 21& 20 & 21&19  \\
			\hline
		\end{tabular}
		\label{T:SRHenOS}
	\end{table}

	$\cnh$ on the HeNOS bench was estimated using a SLODAR method~\cite{Wilson2002} based on WFS cross-correlation, with a 1-km altitude resolution and a measurement precision up to 10\%. Consequently, we expect FPP to retrieve a profile very close to the reference given in Tab.~\ref{T:setupHENOS} \textcolor{black}{within the measurements accuracy given by the WFS cross-correlation method.}
	
\begin{figure}
		\centering
		\includegraphics[width=16cm]{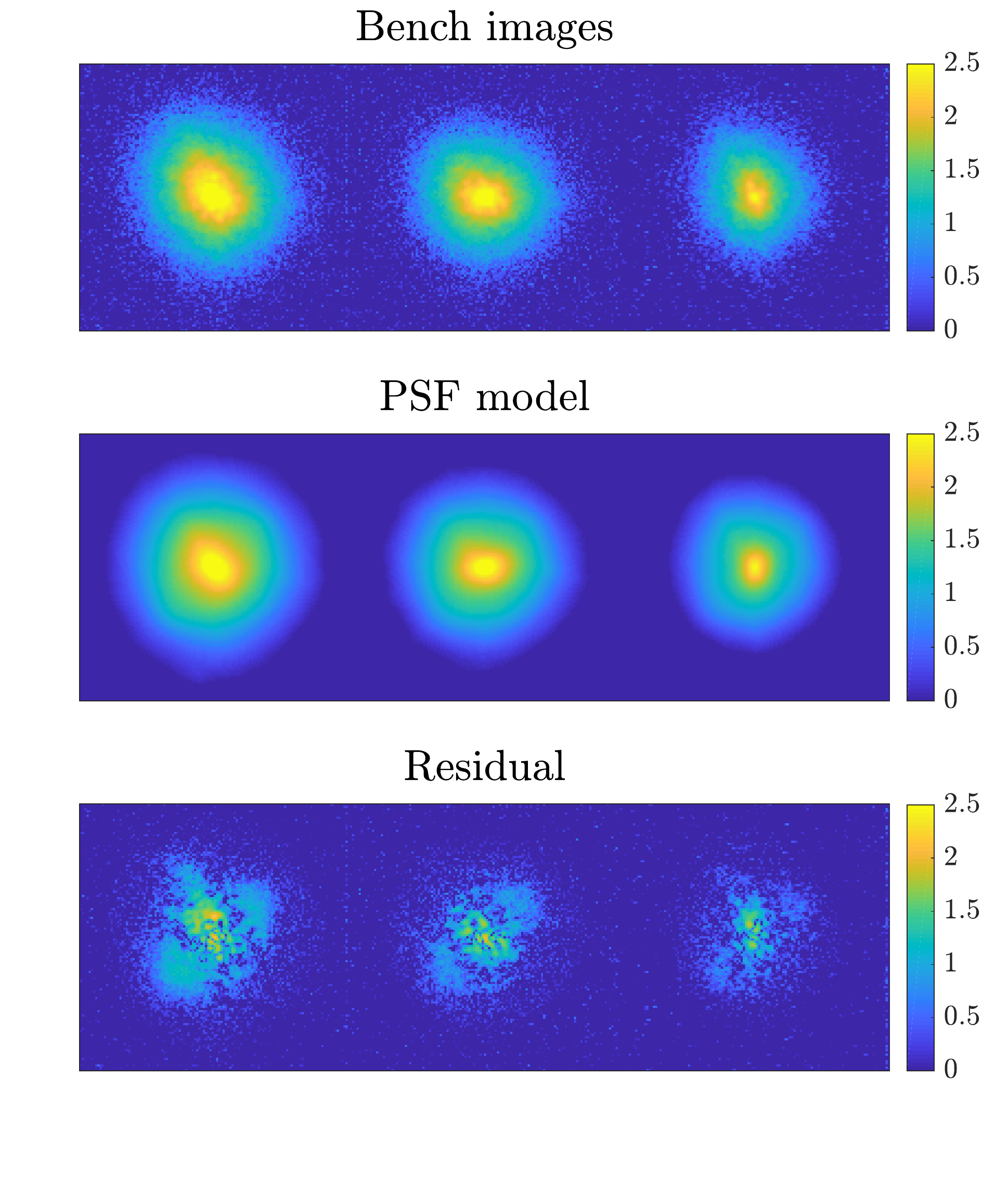}			
		\caption{\small (\textbf{1st line}) off-axis HeNOS PSFs (\textbf{2nd line}) best-fitted PSFs model using FPP (\textbf{3rd line}) fitting residual. Colorbars give the PSF log scale intensity in digital numbers.}
		\label{F:comparPSF}
	\end{figure}

	We have employed FPP to retrieve both weights and heights of three layers by handling one up to three stars, in providing the reference profile as the initial guess. We illustrate in Fig.~\ref{F:comparPSF} the three off-axis PSFs derived using the 3 PSFs-based FPP-outputs profile compared to the observations, that illustrates that our model produces a satisfactory anisoplanatism pattern which matches bench images.
	
\begin{figure}
		\centering	\includegraphics[width=16cm]{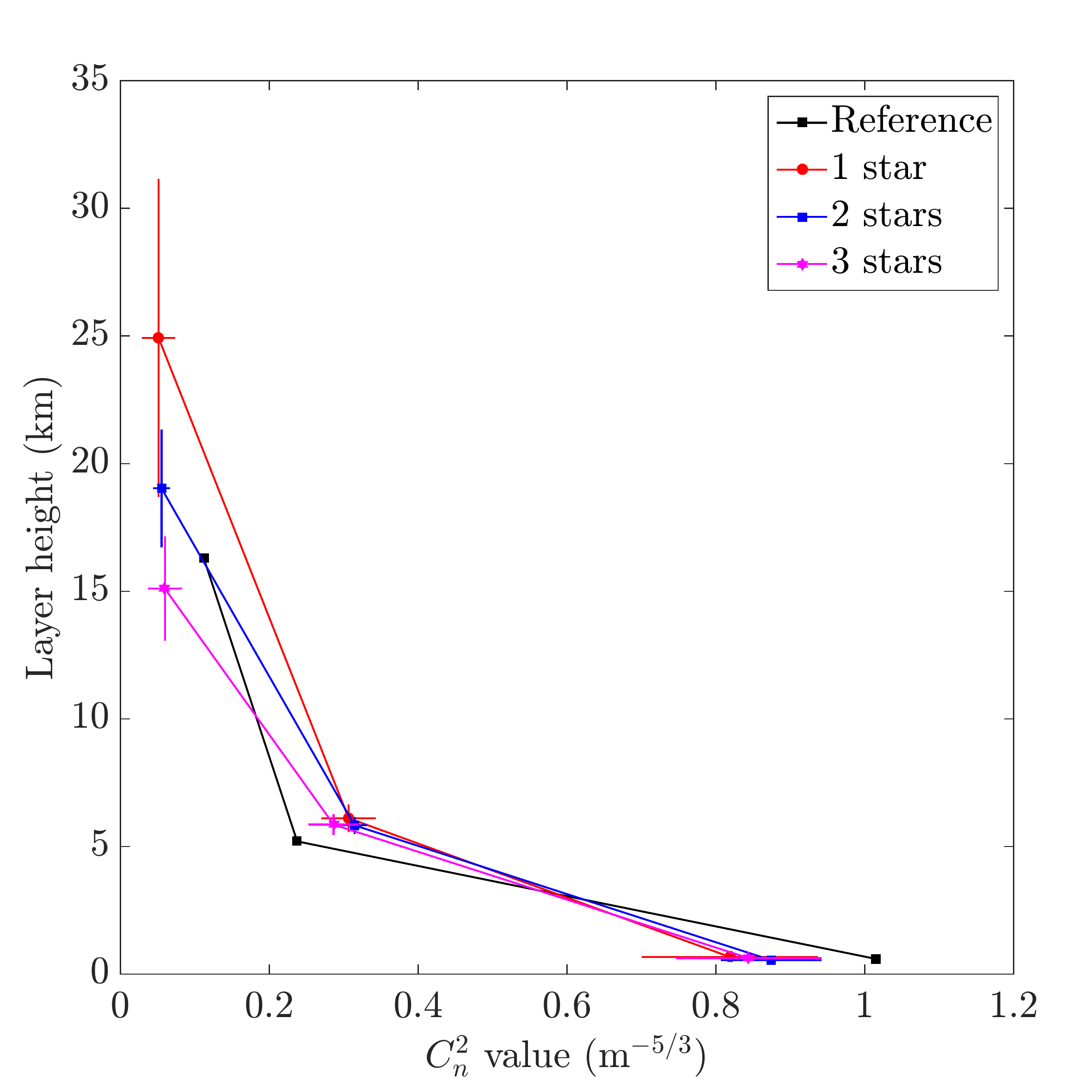}
		\caption{\small Retrieved $\cnh$ profile compared to reference value given in Tab.~\ref{T:setupHENOS}.}
		\label{F:comparAlt}
	\end{figure}

	We provide in Fig.~\ref{F:comparAlt} the averaged $\cnh$ profile over all PSF pairs configurations compared to the reference. Errors bars at 1-$\sigma$ are deduced from a quadratic mean on fitting residual given by the minimisation procedure and averaged out over PSFs pairs~(1/2 PSFs cases).\\
	
	First of all, FPP retrieval depends on the number of stars; we see that the single-PSF case does not lead to a successful retrieval when looking at the highest layer \textcolor{black}{that is strongly overestimated from what we expect}. It is confirmed in Tab.~\ref{T:fvuHenOS} that shows that accuracy on PSF characteristics are worse for the single-star case than using the reference profile as our PSF model inputs. \textcolor{black}{We mention the Fraction of Variance Unexplained (FVU) that derives from the reconstruction residual integrated over all the image, and normalised by the bench mean-free PSF integral as explicated in \cite{BeltramoMartin2018}.}
    
    This is explained by the presence of effects that are not included into the model so far and not or weakly spatially correlated over the three off-axis PSFs, as field-dependent static aberrations for instance. These latter modify the PSF spatial intensity distribution as anisoplanatism does, which may confuse the FPP, especially on a single star.     
    A first lever we have to mitigate unmodelled static phase in the PSF model is to calibrate them as it was done for Keck II~\cite{Witzel2016}. Furthermore, their influence onto the retrieval process is diminished by simply collecting more PSFs in the field as it is illustrated on present results. Contrary to anisoplanatism effect, field-dependent features decorrelates spatially and do not necessarily mimic the elongated anisoplanatism pattern onto the PSF.\\
         
	Identification starts providing a better PSF characterisation starting from two collected PSFs. For the three PSFs-case, we get a $\cnh$ estimation that complies with the 10\%-level precision and 1~km altitude resolution of the WFS-based measurements. Tab.~\ref{T:fvuHenOS} confirms this retrieved set of $\cnh$ values help modelling more accurately the off-axis PSFs. 
	The difference we see may also be introduced by either anisoplanatism model error - a wrong stretch factor would make the equivalent asterism different for instance - or differential conjugation altitude of sources and science camera that may slightly impact the real $\cnh$ as seen in the focal plane compared to WFS-based identification.\\
	
	Fig.~\ref{F:comparAlt} gives the evidence that the retrieved seeing is quite stable and close to the reference value at 10\%-level. On top of that, we notice that only the highest layer altitude estimation is sensitive to the number of stars. This layer contributes the most to spread out the PSF and the FPP tunes the corresponding altitude height to reproduce the FWHM more faithfully.
		    
	The next step of this work is to go further in this identification process by collecting more data in closed-loop on different LGS for increasing the number of observations. We will be able to pinpoint whether this difference is a real physical effect or just a limitation on our system description. Shifting the real position of phase screens will allow to have insights on the FPP altitude resolution as well.
	
	\begin{table}
		\caption{Percentage accuracy on outputs retrieved by FPP with regard to the number of PSF. For 1 and 2 stars cases, we have averaged out over all combination of PSF pairs. Error bars are given at 1-$\sigma$. FVU refers to the fraction of variance unexplained defined in \cite{BeltramoMartin2018}. }
		\centering
		\begin{tabular}{|c|c|c|c|c|}
			\hline
			& 1 star & 2 stars & 3 stars & Reference $\cnh$ \\
			\hline
			FVU & 3.5 & 3.1 & 3.0 & 3.2\\
			\hline
			SR  & 12.0 & 6.7 & 6.5 & 9.9\\		
			\hline
			FWHM & 12.4 & 11.8 & 11.6 & 11.8\\		
			\hline
		\end{tabular}
		\label{T:fvuHenOS}
	\end{table}

	\section{Conclusion}
	\label{S:conclusions}
	
	We present in this paper the focal plane profiling as a $\cnh$ retrieval method that relies on partially compensated AO images affected by anisoplanatism. It performs a non-linear least-square minimisation of a PSF model over observations and provide both a PSF model across the field and the $\cnh$ profile. To mitigate noise propagation and the sensitivity to unmodelled aberrations, such as field-dependent static aberrations, we must collect several PSF from the field; we show for the NIRC2 imaging camera at Keck II that the FPP can retrieve both atmosphere and PSF characteristics, photometry at 1\%-level accuracy and 50$\mu$as astrometry if we get $n_\text{psf}$ PSFs of magnitude given by $H=14 + 2.5\times\log(n_\text{psf}\times T_\text{exp}/30)$, with a breakdown to H=15.5 mag when picking off 4 stars in the field.
	
	We have deployed this approach on the HeNOS testbench where $\cnh$ values are measured from WFS cross-correlation. Thanks to FPP, we retrieved a profile that complies with WFS-based measurements when using three stars distributed over 4.5" with $\theta_0 = 0.854"$. We demonstrate that collecting more stars allows to mitigate model errors such as field-dependent static aberrations for instance.\\
	
	We focused in this paper for the $\cnh$ profiling in the purpose of assessing the reliability and limitations of this method. Our next work will consist in deploying both FPP and PSF-reconstruction technique to asses what are the potential gains on crowded fields observations, such as the Galactic center with Keck and the ELT with MICADO.   
    Furthermore, we will investigate to extend the FPP to tomographic systems for improving the $\cnh$ profiling, especially in the purpose to deploy such an approach to the multi-conjugated system MAORY coupled with MICADO, or the laser-tomographic mode of HARMONI and METIS.\\
	
\section*{Acknowledgements}
    
	This work was supported by the A*MIDEX project (no. ANR-11-IDEX-0001-02) funded by the "Investissements d'Avenir" French Government programme, managed by the French National Research Agency (ANR).

\bibliographystyle{plain} 
\bibliography{../biblioLolo}

\end{document}